\documentclass{article}

\usepackage{graphicx}
\usepackage{glossaries}
\usepackage{booktabs}
\usepackage[numbers]{natbib}
\usepackage{tikz}
\usepackage{pgfplots}
\usepackage{subcaption}
\usepackage{array}
\usepackage{attachfile}
\usepackage[capitalise,noabbrev,sort]{cleveref}
\usepackage{authblk}
\usepackage{multirow}
\usepackage{makecell}
\usepackage{amssymb}
\usepackage{makecell} 
\setcellgapes{5pt}

\usetikzlibrary{positioning}
\usetikzlibrary{shapes,arrows, spy}
\usetikzlibrary{fit}
\usetikzlibrary{arrows.meta}

\newcolumntype{L}[1]{>{\raggedright\let\newline\\\arraybackslash\hspace{0pt}}m{#1}}
\newcolumntype{C}[1]{>{\centering\let\newline\\\arraybackslash\hspace{0pt}}m{#1}}
\newcolumntype{R}[1]{>{\raggedleft\let\newline\\\arraybackslash\hspace{0pt}}m{#1}}

\newacronym{CNN}{CNN}{convolutional neural network}
\newacronym{MRI}{MRI}{magnetic resonance imaging}
\newacronym{WHO}{WHO}{World Health Organization}
\newacronym{T1}{T1w}{T1-weighted}
\newacronym{T1C}{T1wC}{post-contrast T1-weighted}
\newacronym{T2}{T2w}{T2-weighted}
\newacronym{FLAIR}{T2w-FLAIR}{T2-weighted fluid attenuated inversion recovery}
\newacronym{IDH}{IDH}{isocitrate dehydrogenase}
\newacronym{LGG}{LGG}{low grade glioma}
\newacronym{HGG}{HGG}{high grade glioma}
\newacronym{MGMT}{MGMT}{O$^6$-methylguanine-methyltransferase}
\newacronym{TERT}{TERT}{telomerase reverse transcriptase}
\newacronym{TCIA}{TCIA}{The Cancer Imaging Archive}
\newacronym{AUC}{AUC}{area under receiver operating characteristic curve}
\newacronym{GBM}{GBM}{glioblastoma}

\newacronym{DWI}{DWI}{diffusion-weighted imaging}
\newacronym{PWI}{PWI}{perfusion-weighted imaging}

\newacronym{BRATS}{BraTS}{Brain Tumor Segmentation challenge}
\newacronym{NIFTI}{NIfTI}{Neuroimaging Informatics Technology Initiative}

\newacronym{TP}{TP}{true positive}
\newacronym{TN}{TN}{true negative}
\newacronym{FP}{FP}{false positive}
\newacronym{FN}{FN}{false negative}

\newacronym{EMC}{EMC}{Erasmus MC}
\newacronym{HMC}{HMC}{Haaglanden Medical Center}
\newacronym{AMC}{AUMC}{Amsterdam UMC}
\newacronym{UMCU}{UMCU}{University Medical Center Utrecht}

\newacronym{CPTAC}{CPTAC-GBM}{Clinical Proteomic Tumor Analysis Consortium Glioblastoma Multiforme}
\newacronym{IVY}{Ivy GAP}{Ivy Glioblastoma Atlas Project}
\newacronym{BTP}{Brain-Tumor-Progression}{Brain-Tumor-Progression}
\newacronym{REMBRANDT}{REMBRANDT}{Repository of Molecular Brain Neoplasia Data}
\newacronym{TCGALGG}{TCGA-LGG}{TCGA-LGG}
\newacronym{TCGAGBM}{TCGA-GBM}{TCGA-GBM}

\glsunset{TCGALGG}
\glsunset{TCGAGBM}
\glsunset{BTP}

\newacronym{ROC}{ROC}{receiver operating characteristic}

\newcommand{\acrcaption}[1]{\textbf{\acrshort{#1}}: \acrlong{#1}}

\title{\textbf{WHO 2016 subtyping and automated segmentation of glioma using multi-task deep learning}}

\author[1,$\dagger$]{Sebastian R. van der Voort}
\author[2,3,$\dagger$]{Fatih Incekara}
\author[4]{Maarten MJ Wijnenga}
\author[2]{Georgios Kapsas}
\author[2]{Renske Gahrmann}
\author[3]{Joost W Schouten}
\author[5]{Rishi Nandoe Tewarie}
\author[6]{Geert J Lycklama}
\author[7]{Philip C De Witt Hamer}
\author[7]{Roelant S Eijgelaar}
\author[4]{Pim J French}
\author[8]{Hendrikus J Dubbink}
\author[3]{Arnaud JPE Vincent}
\author[1,9]{Wiro J Niessen}
\author[4]{Martin J van den Bent}
\author[2,$\dagger\dagger$]{Marion Smits}
\author[1,$\dagger\dagger$,*]{Stefan Klein}

\affil[1]{Biomedical Imaging Group Rotterdam, Department of Radiology and Nuclear Medicine, Erasmus MC University Medical Centre Rotterdam, Rotterdam, the Netherlands}
\affil[2]{Department of Radiology and Nuclear Medicine, Erasmus MC University Medical Centre Rotterdam, Rotterdam, the Netherlands}
\affil[3]{Department of Neurosurgery, Brain Tumor Center, Erasmus MC University Medical Centre Rotterdam, Rotterdam, the Netherlands}
\affil[4]{Department of Neurology, Brain Tumor Center, Erasmus MC Cancer Institute, Rotterdam, the Netherlands}
\affil[5]{Department of Neurosurgery, Haaglanden Medical Center, the Hague, the Netherlands}
\affil[6]{Department of Radiology, Haaglanden Medical Center, the Hague, the Netherlands}
\affil[8]{Department of Pathology, Brain Tumor Center at Erasmus MC Cancer Institute, Rotterdam, the Netherlands}
\affil[9]{Imaging Physics, Faculty of Applied Sciences, Delft University of Technology, Delft, the Netherlands}
\affil[7]{Department of Neurosurgery, Cancer Center Amsterdam, Brain Tumor Center, Amsterdam UMC, Amsterdam, Netherlands}
\affil[$\dagger$]{These authors contributed equally}
\affil[$\dagger\dagger$]{These authors contributed equally}
\affil[*]{Corresponding author; s.klein@erasmusmc.nl}
\date{}

\begin{document}
\maketitle

\newpage
\begin{abstract}
    Accurate characterization of glioma is crucial for clinical decision making.
    A delineation of the tumor is also desirable in the initial decision stages but is a time-consuming task.
    Leveraging the latest GPU capabilities, we developed a single multi-task convolutional neural network that uses the full 3D, structural, pre-operative MRI scans to can predict the IDH mutation status, the 1p/19q co-deletion status, and the grade of a tumor, while simultaneously segmenting the tumor.
    We trained our method using the largest, most diverse patient cohort to date containing 1508 glioma patients from 16 institutes.
    We tested our method on an independent dataset of 240  patients from 13 different institutes, and achieved an IDH-AUC of 0.90, 1p/19q-AUC of 0.85, grade-AUC of 0.81, and a mean whole tumor DICE score of 0.84.
    Thus, our method non-invasively predicts multiple, clinically relevant parameters and generalizes well to the broader clinical population.
\end{abstract}

\section{Introduction}
Glioma is the most common primary brain tumor and is one of the deadliest forms of cancer \cite{office2019cancer}.
Differences in survival and treatment response of glioma are attributed to their genetic and histological features, specifically the \gls{IDH} mutation status, the 1p/19q co-deletion status and the tumor grade \cite{dubbink2015molecular, eckel2015gliomagroups}.
Therefore, in 2016 the \gls{WHO} updated its brain tumor classification, categorizing glioma based on these genetic and histological features \cite{louis20162016}.
In current clinical practice, these features are determined from tumor tissue.
While this is not an issue in patients in whom the tumor can be resected, this is problematic when resection can not safely be performed.
In these instances,  surgical biopsy is performed with the sole purpose of obtaining tissue for diagnosis, which, although relatively safe, is not without risk \cite{chen2009biopsy, jackson2001biopsylimitations}.
Therefore, there has been an increasing interest in complementary non-invasive alternatives that can provide the genetic and histological information used in the \gls{WHO} 2016 categorization \cite{zhou2018radiomicsbrain, bi2019AIcancer}.

\Gls{MRI} has been proposed as a possible candidate because of its non-invasive nature and its current place in routine clinical care \cite{chaddad2019radiomicsglioblastoma}.
Research has shown that certain \gls{MRI} features, such as the tumor heterogeneity, correlate with the genetic and histological features of glioma \cite{smits2016imaging, delfanti2017imagingcorrelates}.
This notion has popularized, in addition to already popular applications such as tumor segmentation, the use of machine learning methods for the prediction of genetic and histological features, known as radiomics \cite{gore2020review, aerts2014radiomics, chichate2020review}.
Although a plethora of such methods now exist, they have found little translation to the clinic \cite{gore2020review}.

An often discussed challenge for the adoption of machine learning methods in clinical practice is the lack of standardization, resulting in heterogeneity of patient populations, imaging protocols, and scan quality \cite{gillies2016radiomics, thrall2018AIradiology}.
Since machine learning methods are prone to overfitting, this heterogeneity questions the validity of such methods in a broader patient population \cite{thrall2018AIradiology}.
Furthermore, it has been noted that most current research concerns narrow task-specific methods that lack the context between different related tasks, which might restrict the performance of these methods \cite{hosny2018AI}.

An important technical limitation when using deep learning methods is the limited GPU memory, which restricts the size of models that can be trained \cite{kopuklu2019cnn}.
This is a problem especially for clinical data, which is often 3D, requiring even more memory than the commonly used 2D networks.
This further limits the size of these models resulting in shallower models, and the use of patches of a scan instead of using the full 3D scan as an input, which limits the amount of context these methods can extract from the scans.

\begin{figure}
    \includegraphics[width=\textwidth]{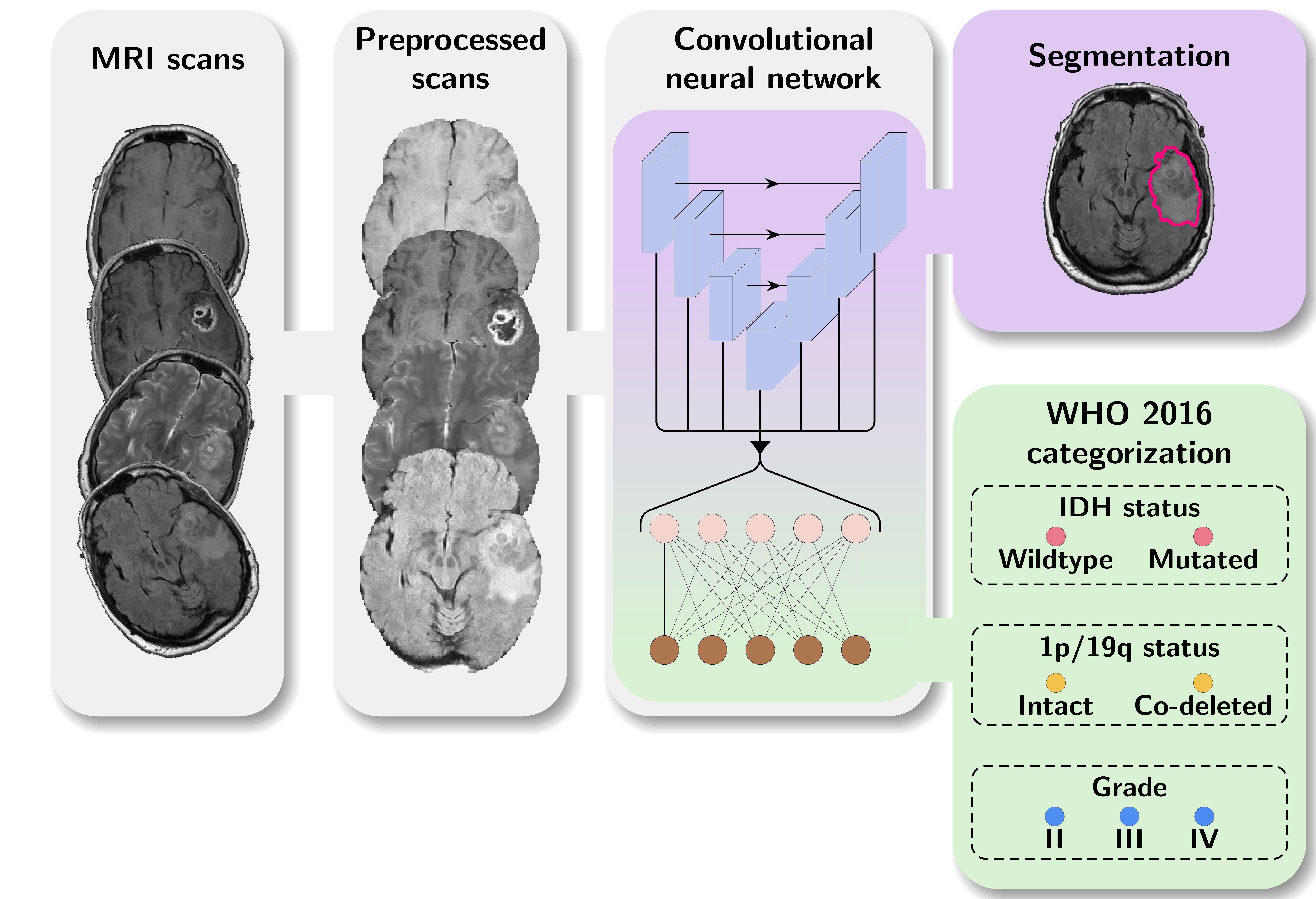}
    \caption{Overview of our method.
    Pre-, and post-contrast \gls{T1}, \gls{T2} and \gls{FLAIR} scans are used as an input.
    The scans are registered to an atlas, bias field corrected, skull stripped, and normalized before being passed through our convolutional neural network.
    One branch of the network segments the tumor, while at the same time the features are combined to predict the IDH status, 1p/19q status, and grade of the tumor.
    }\label{fig:pipeline}
    \end{figure}

Here, we present a new method that addresses the above problems.
Our method consists of a single, multi-task \gls{CNN} that can predict the \gls{IDH} mutation status, the 1p/19q co-deletion status, and the grade (grade II/III/IV) of a tumor, while also simultaneously segmenting the tumor, see \cref{fig:pipeline}.
To the best of our knowledge, this is the first method that provides all of this information at the same time, allowing clinical experts to derive the \gls{WHO} category from the individually predicted genetic and histological features, while also allowing them to consider or disregard specific predictions as they deem fit.
Exploiting the capabilities of the latest GPUs, optimizing our implementation to reduce the memory footprint, and using distributed multi-GPU training, we were able to train a model that uses the full 3D scan as an input.
We trained our method using the largest, most diverse patient cohort to date, with 1508 patients included from 16 different institutes.
To ensure the broad applicability of our method, we used minimal inclusion criteria, only requiring the four most commonly used \gls{MRI} sequences: pre- and post-contrast \gls{T1}, \gls{T2}, and \gls{FLAIR} \cite{thust2018gliomaimaging, fresychlag2018imaging}.
No constraints were placed on the patients' clinical characteristics, such as the tumor grade, or the radiological characteristics of scans, such as the scan quality.
In this way, our method could capture the heterogeneity that is naturally present in clinical data.
We tested our method on an independent dataset of 240 patients from 13 different institutes, to evaluate the true generalizability of our method.
Our results show that we can predict multiple clinical features of glioma from \gls{MRI} scans in a diverse patient population.

\newpage

\section{Results}
\subsection{Patient characteristics}
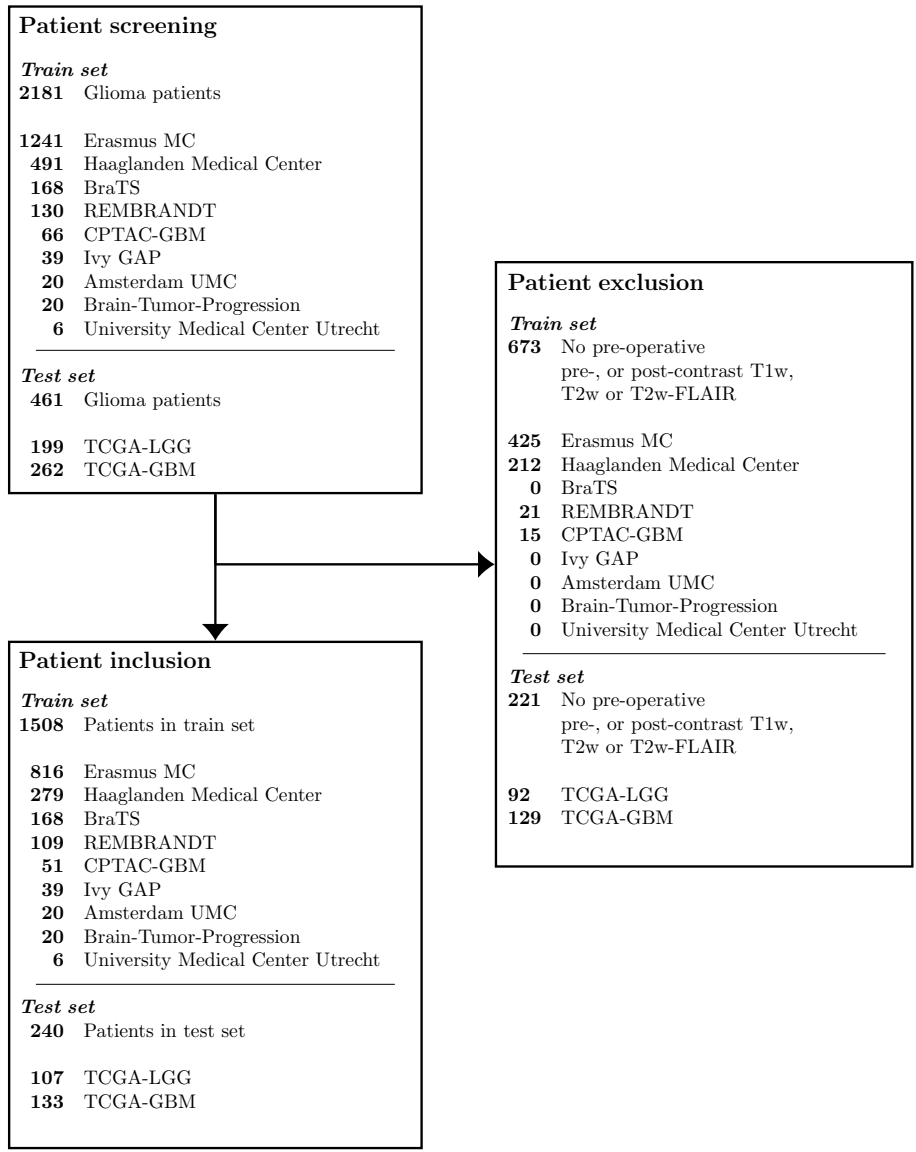
\begin{figure}
\centering
\resizebox{\textwidth}{!}{\begin{tikzpicture}[auto,
    block_center/.style ={rectangle, draw=black, very thick, fill=white,
      text width=25em, text centered,
      minimum height=4em},
    algorithm_center/.style ={circle, draw=black, very thick, fill=white,
    text width = 7em, text ragged, minimum height=1em, inner sep=3pt, text centered},
    block_left/.style ={rectangle, draw=black, very thick, fill=white,
      text width=20em, text ragged, minimum height=2em, inner sep=5pt},
    exclusion_block_left/.style ={rectangle, draw=black, very thick, fill=white,
    text width=20em, text ragged, minimum height=4em, inner sep=6pt},
    block_test/.style ={rectangle, draw=black, thick, fill=white,
    text width=10em, text ragged, minimum height=4em, inner sep=6pt},
      line/.style ={draw, very thick, -{Latex[length=3mm,width=5mm]}, shorten >=0pt}]

     \node [block_left] (identified) {%
     {\large\textbf{Patient screening}}
    \\
     \begin{tabbing}
      \textit{\textbf{Train set}}\\
      \=\textbf{2}\=\textbf{1}\=\textbf{8}\=\textbf{1}\hspace{1em}\=Glioma patients\\
      \\
      \>\textbf{1241}\>\>\>\>\acrlong{EMC}\\
      \>\>\textbf{491}\>\>\>\acrlong{HMC}\\
      \>\>\textbf{168}\>\>\>\acrshort{BRATS}\\
      \>\>\textbf{130}\>\>\>\acrshort{REMBRANDT}\\
      \>\>\>\textbf{66}\>\>\acrshort{CPTAC}\\
      \>\>\>\textbf{39}\>\>\acrshort{IVY}\\
      \>\>\>\textbf{20}\>\>\acrlong{AMC}\\
      \>\>\>\textbf{20}\>\>\acrshort{BTP}\\
      \>\>\>\>\textbf{6}\>\acrlong{UMCU}\\
      \\

    \textit{\textbf{Test set}}\\
      \>\>\textbf{461}\>\>\>Glioma patients\\
      \\
      \>\>\textbf{199}\>\>\>\acrshort{TCGALGG}\\
      \>\>\textbf{262}\>\>\>\acrshort{TCGAGBM}
    \end{tabbing}
    };

    \draw ([yshift=-1.8cm, xshift=5mm] identified.west) -- ([yshift=-1.8cm, xshift=-5mm] identified.east);

     \node [block_left, below=7cm of identified.center] (included) {
      {\large\textbf{Patient inclusion}}
     \begin{tabbing}
      \textit{\textbf{Train set}}\\
     \=\textbf{1}\=\textbf{5}\=\textbf{0}\=\textbf{8}\hspace{1em}\=Patients in train set\\
     \\
     \>\>\textbf{816}\>\>\>\acrlong{EMC}\\
     \>\>\textbf{279}\>\>\>\acrlong{HMC}\\
     \>\>\textbf{168}\>\>\>\acrshort{BRATS}\\
     \>\>\textbf{109}\>\>\>\acrshort{REMBRANDT}\\
     \>\>\>\textbf{51}\>\>\acrshort{CPTAC}\\
     \>\>\>\textbf{39}\>\>\acrshort{IVY}\\
     \>\>\>\textbf{20}\>\>\acrlong{AMC}\\
     \>\>\>\textbf{20}\>\>\acrshort{BTP}\\
     \>\>\>\>\textbf{6}\>\acrlong{UMCU}\\
     \\
     \textit{\textbf{Test set}}\\
     \>\>\textbf{240}\>\>\>Patients in test set\\
     \\
     \>\>\textbf{107}\>\>\>\acrshort{TCGALGG}\\
     \>\>\textbf{133}\>\>\>\acrshort{TCGAGBM}\\
     \end{tabbing}};

     \draw ([yshift=-1.6cm, xshift=5mm] included.west) -- ([yshift=-1.6cm, xshift=-5mm] included.east);

     \node [exclusion_block_left, below right=0.2cm and 5cm of identified.center] (excluded) {%
     {\large\textbf{Patient exclusion}}
     \begin{tabbing}
      \textit{\textbf{Train set}}\\
      \=\textbf{6}\=\textbf{7}\=\textbf{3}\hspace{1em}\=No pre-operative \\
                   \>\>\>\> pre-, or post-contrast \acrshort{T1},\\
                   \>\>\>\> \acrshort{T2} or \acrshort{FLAIR}\\
      \\
      \>\textbf{425}\>\>\>\acrlong{EMC}\\
      \>\textbf{212}\>\>\>\acrlong{HMC}\\
      \>\>\>\textbf{0}\>\acrshort{BRATS}\\
      \>\>\textbf{21}\>\>\acrshort{REMBRANDT}\\
      \>\>\textbf{15}\>\>\acrshort{CPTAC}\\
      \>\>\>\textbf{0}\>\acrshort{IVY}\\
      \>\>\>\textbf{0}\>\acrlong{AMC}\\
      \>\>\>\textbf{0}\>\acrshort{BTP}\\
      \>\>\>\textbf{0}\>\acrlong{UMCU}\\
      \\
      \textit{\textbf{Test set}}\\
      \>\textbf{221}\>\>\>No pre-operative \\
                   \>\>\>\> pre-, or post-contrast \acrshort{T1},\\
                   \>\>\>\> \acrshort{T2} or \acrshort{FLAIR}\\
      \\
      \>\textbf{92}\>\>\>\acrshort{TCGALGG}\\
      \>\textbf{129}\>\>\>\acrshort{TCGAGBM}\\

    \end{tabbing}
     };

     \draw ([yshift=-1.6cm, xshift=5mm] excluded.west) -- ([yshift=-1.6cm, xshift=-5mm] excluded.east);





    \begin{scope}[every path/.style=line]
        \path (identified.south)   -- (included);
        \path (identified)   |- (excluded);


    \end{scope}

  \end{tikzpicture}}
\caption{Inclusion flowchart of the train set and test set.}\label{fig:inclusion_flowchart}
\end{figure}

We included a total of 1748 patients in our study, 1508 as a train set and 240 as an independent test set.
The patients in the train set originated from nine different data collections and 16 different institutes, and the test set was collected from two different data collections and 13 different institutes.
\cref{tab:patient_char} provides a full overview of the patient characteristics in the train and test set, and \cref{fig:inclusion_flowchart} shows the inclusion flowchart and the distribution of the patients over the different data collections in the train set and test set.

\begin{table}[htbp]
\centering
\caption{Patient characteristics for the train set and test set.}\label{tab:patient_char}
\begin{tabular}{l r r C{0.5cm} r r}
\toprule
&\multicolumn{2}{c}{\textbf{Train set}} && \multicolumn{2}{c}{\textbf{Test set}}\\
&N&\% && N &\%\\
\midrule
Patients & 1508 &  && 240\\

IDH status \\
\hspace{1em}Mutated & 226 & 15.0 && 88 &36.7\\
\hspace{1em}Wildtype & 440 & 29.2 && 129&53.7\\
\hspace{1em}Unknown & 842 & 55.8 && 23 & 9.6\\

1p/19q co-deletion status\\
\hspace{1em}Co-deleted & 103 & 6.8 && 26 & 10.8\\
\hspace{1em}Intact & 337 & 22.4 &&207 & 86.3\\
\hspace{1em}Unknown & 1068 & 70.8 && 7 &2.9\\

Grade\\
\hspace{1em}II& 230 & 15.3 && 47 &19.6\\
\hspace{1em}III& 114 & 7.6 && 59 & 24.6\\
\hspace{1em}IV& 830 & 55.0 && 132 & 55.0\\
\hspace{1em}Unknown& 334 & 22.1 && 2 & 0.8\\

WHO 2016 categorization\\
\hspace{1em}Oligodendroglioma & 96 &6.4 && 26 &10.8\\
\hspace{1em}Astrocytoma, IDH wildtype & 31 & 2.1 && 22 & 9.2\\
\hspace{1em}Astrocytoma, IDH mutated & 98 & 6.4 && 57 & 23.7\\
\hspace{1em}GBM, IDH wildtype & 331 & 21.9 && 106 &44.2\\
\hspace{1em}GBM, IDH mutated & 16 & 1.1 && 5 & 2.1\\
\hspace{1em}Unknown & 936 & 62.1 && 24 & 10.0\\

Segmentation\\
\hspace{1em}Manual & 716 & 47.5 && 240 & 100\\
\hspace{1em}Automatic & 792 & 52.5 && 0 & 0\\

\bottomrule
\end{tabular}

{\small \raggedright \acrcaption{IDH}, \acrcaption{WHO}, \acrcaption{GBM} \par}
\end{table}

\subsection{Algorithm performance}

We used 15\% of the train set as a validation set and selected the model parameters that achieved the best performance on this validation set, where the model achieved an \gls{AUC} of 0.88 for the \gls{IDH} mutation status prediction, an \gls{AUC} of 0.76 for the 1p/19q co-deletion prediction, an \gls{AUC} of 0.75 for the grade prediction, and a mean segmentation DICE score of 0.81.
The selected model parameters are shown in \cref{app:hyperparameter_tuning}.
We then trained a model using these parameters and the full train set, and evaluated it on the independent test set.

\begin{figure}[htbp]
\centering
\includegraphics[width=\textwidth]{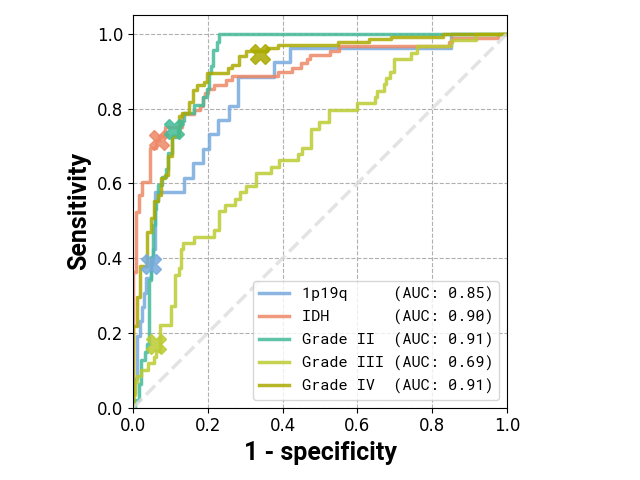}
\caption{\Acrfull{ROC}-curves of the genetic and histological features, evaluated on the test set. The crosses indicate the location of the decision threshold for the reported accuracy, sensitivity, and specificity.}\label{fig:roc_curves}
\end{figure}

For the genetic and histological feature predictions, we achieved an AUC of 0.90 for the \gls{IDH} mutation status prediction, an \gls{AUC} of 0.85 for the 1p/19q co-deletion prediction, and an \gls{AUC} of 0.81 for the grade prediction, in the test set.
The full results are shown in \cref{tab:cnn_results}, with the corresponding \gls{ROC}-curves in \cref{fig:roc_curves}.
\cref{tab:cnn_results} also shows the results in (clinically relevant) subgroups of patients.
This shows that we achieved an \gls{IDH}-AUC of 0.81 in \gls{LGG} (grade II/III), an \gls{IDH}-AUC of 0.64 in \gls{HGG} (grade IV), and a 1p/19q-AUC of 0.73 in \gls{LGG}.
When only predicting \gls{LGG} vs. \gls{HGG} instead of predicting the individual grades, we achieved an AUC of 0.91.
In \cref{app:confusion_matrix} we provide confusion matrices for the \gls{IDH}, 1p/19q, and grade predictions, as well as a confusion matrix for the final \gls{WHO} 2016 subtype, which shows that only one patient was predicted as a non-existing \gls{WHO} 2016 subtype.
In  \cref{app:individual_pred} we provide the individual predictions and ground truth labels for all patients in the test set to allow for the calculation of additional metrics.

\begin{figure}[htbp]
    \centering
    \includegraphics[width=\textwidth]{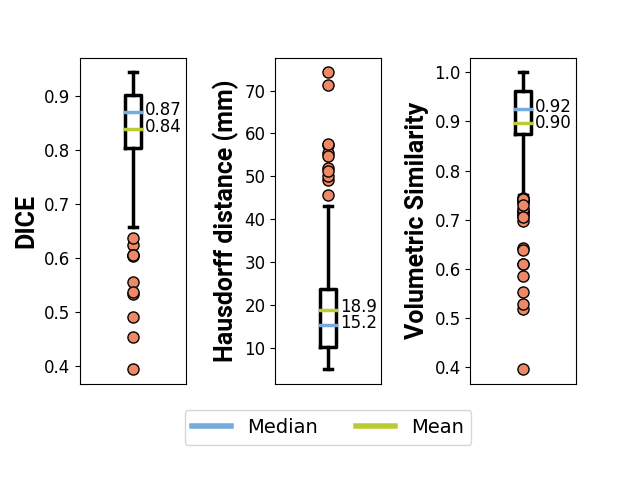}
    \caption{DICE scores, Hausdorff distances, and volumetric similarity coefficients for all patients in the test set.
    The DICE score is a measure of the overlap between the ground truth and predicted segmentation (where 1 indicates perfect overlap).
    The Hausdorff distance is a measure of the agreement between the boundaries of the ground truth and predicted segmentation (lower is better).
    The volumetric similarity coefficient is a measure of the agreement in volume (where 1 indicates perfect agreement).}\label{fig:segmentation_results}
\end{figure}

For the automatic segmentation, we achieved a mean DICE score of 0.84, a mean Hausdorff distance of 18.9 mm, and a mean volumetric similarity coefficient of 0.90.
\cref{fig:segmentation_results} shows boxplots of the DICE scores, Hausdorff distances, and volumetric similarity coefficients for the different patients in the test set.
In \cref{app:seg_examples_random} we show five patients that were randomly selected from both the \gls{TCGALGG} and \gls{TCGAGBM} data collections, to demonstrate the automatic segmentations made by our method.

\begin{table}[htbp]
    \centering
    \caption{Evaluation results of the final model on the test set.}\label{tab:cnn_results}
    \begin{tabular}{llC{1.7cm}C{1.7cm}C{1.7cm}C{1.7cm}}
        \toprule
        \thead{Patient\\group} & Task &  AUC & Accuracy & Sensitivity & Specificity \\
        \midrule

        All& IDH & 0.90 & 0.84 & 0.72 & 0.93\\
        &1p/19q & 0.85 & 0.89 & 0.39 & 0.95\\
        &Grade (II/III/IV)& 0.81 & 0.71 & N/A & N/A\\
        &Grade II & 0.91 & 0.86 &  0.75 & 0.89\\
        &Grade III & 0.69 & 0.75 & 0.17 & 0.94\\
        &Grade IV & 0.91 & 0.82 & 0.95 & 0.66\\
        &LGG vs HGG & 0.91 & 0.84 & 0.72 & 0.93\\

        \\
        LGG & IDH & 0.81 & 0.74 & 0.73 & 0.77\\
        &1p/19q& 0.73 & 0.76 & 0.39 & 0.89\\

        \\
        HGG & IDH &  0.64 & 0.94 & 0.40 & 0.96\\
        \bottomrule
        \end{tabular}

        {\small \raggedright Abbreviations: \acrcaption{AUC}, \acrcaption{IDH}, \acrcaption{LGG}, \acrcaption{HGG} \par}
\end{table}

\subsection{Model interpretability}
\begin{figure}[htbp]
    \centering
    \begin{subfigure}[b]{\textwidth}
        \centering

        \hfill
        \begin{subfigure}[b]{0.24\textwidth}
        \includegraphics[width=\textwidth]{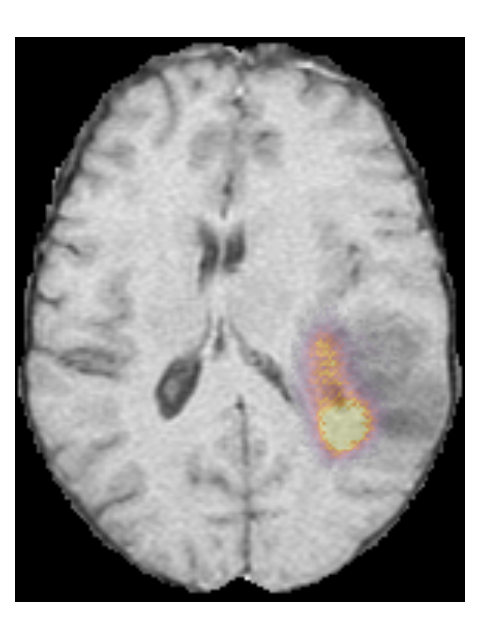}
        \caption*{Pre-contrast \acrshort{T1}}
        \end{subfigure}
        \hfill
        \begin{subfigure}[b]{0.24\textwidth}
        \includegraphics[width=\textwidth]{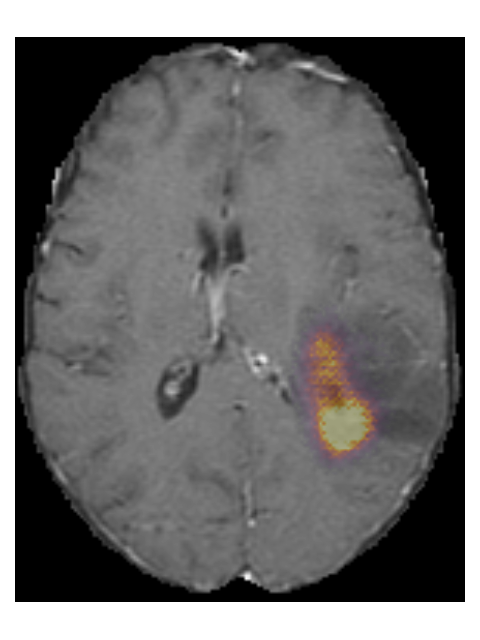}
        \caption*{Post-contrast \acrshort{T1}}
        \end{subfigure}
        \hfill
        \begin{subfigure}[b]{0.24\textwidth}
        \includegraphics[width=\textwidth]{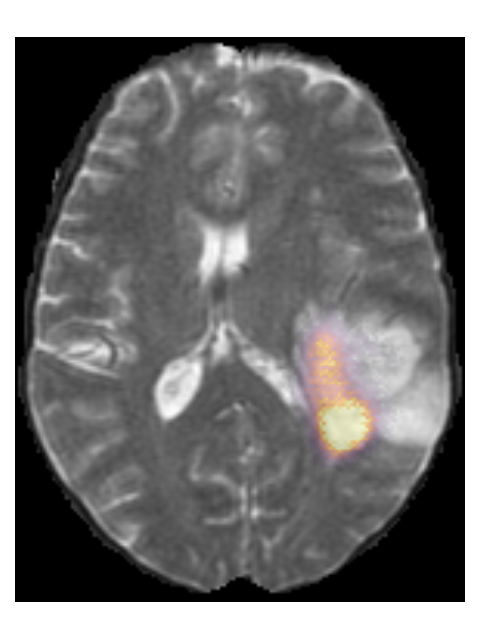}
        \caption*{\acrshort{T2}}
        \end{subfigure}
        \hfill
        \begin{subfigure}[b]{0.24\textwidth}
        \includegraphics[width=\textwidth]{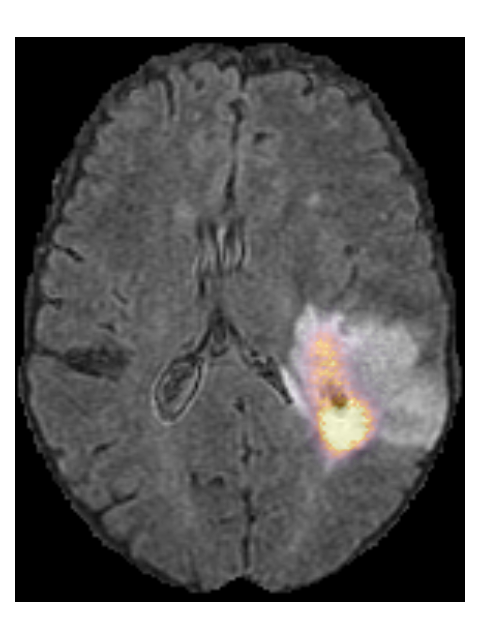}
        \caption*{\acrshort{FLAIR}}
        \end{subfigure}
        \caption{Saliency maps of a low grade glioma patient (TCGA-DU-6400). This is an IDH mutated, 1p/19q co-deleted, grade II tumor.
        The network focuses on a rim of brightness in the \gls{FLAIR} scan.}\label{fig:saliency_LGG}
    \end{subfigure}
    \vskip\baselineskip
    \begin{subfigure}[b]{\textwidth}
        \centering
        \hfill
        \begin{subfigure}[b]{0.24\textwidth}
        \includegraphics[width=\textwidth]{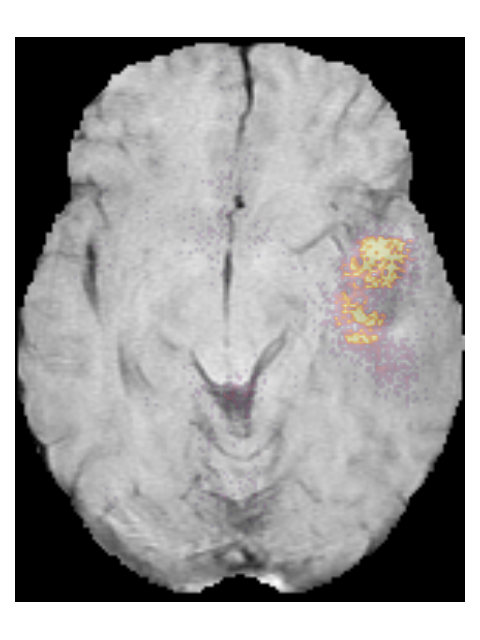}
        \caption*{Pre-contrast \acrshort{T1}}
        \end{subfigure}
        \hfill
        \begin{subfigure}[b]{0.24\textwidth}
        \includegraphics[width=\textwidth]{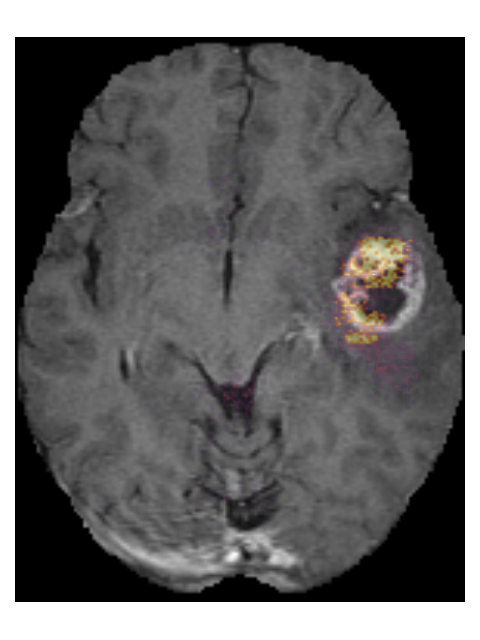}
        \caption*{Post-contrast \acrshort{T1}}
        \end{subfigure}
        \hfill
        \begin{subfigure}[b]{0.24\textwidth}
        \includegraphics[width=\textwidth]{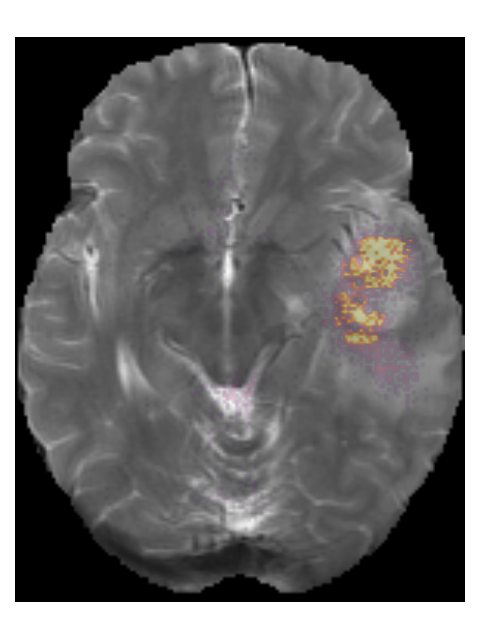}
        \caption*{\acrshort{T2}}
        \end{subfigure}
        \hfill
        \begin{subfigure}[b]{0.24\textwidth}
        \includegraphics[width=\textwidth]{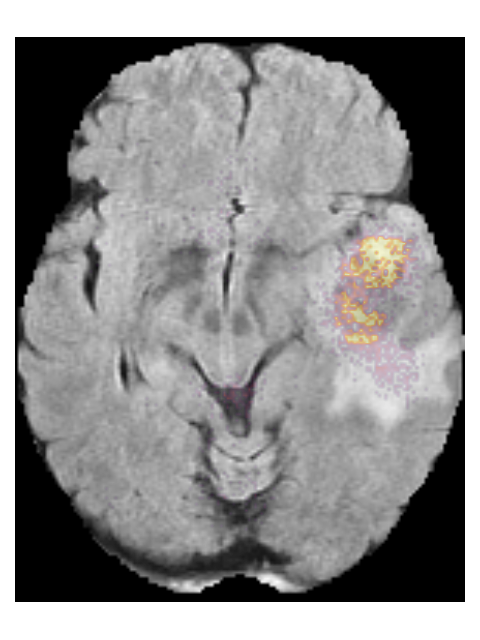}
        \caption*{\acrshort{FLAIR}}
        \end{subfigure}
        \caption{Saliency maps of a high grade glioma patient (TCGA-06-0238). This is an IDH wildtype, grade IV tumor.
        The network focuses on enhancing spots around the necrosis on the post-contrast \gls{T1} scan.}\label{fig:saliency_HGG}
    \end{subfigure}

\caption{Saliency maps of two patients from the test set, showing areas that are relevant for the prediction.}\label{fig:saliency_maps}
\end{figure}

To provide insight into the behavior of our model we created saliency maps, which show which parts of the scans contributed the most to the prediction.
These saliency maps are shown in \cref{fig:saliency_maps} for two example patients from the test set.
It can be seen that for the \gls{LGG} the network focused on a bright rim in the \gls{FLAIR} scan, whereas for the \gls{HGG} it focused on the enhancement in the post-contrast \gls{T1} scan.
To aid further interpretation, we provide visualizations of selected filter outputs in the network in \cref{app:filter_vis}, which also show that the network focuses on the tumor, and these filters seem to recognize specific imaging features such as the contrast enhancement and \gls{FLAIR} brightness.

\subsection{Model robustness}

\begin{figure}[htbp]
    \centering
    \begin{subfigure}[b]{\textwidth}
        \centering
        \hfill
        \includegraphics[width=0.43\textwidth, clip, trim=2.5cm 0cm 2.5cm 0cm]{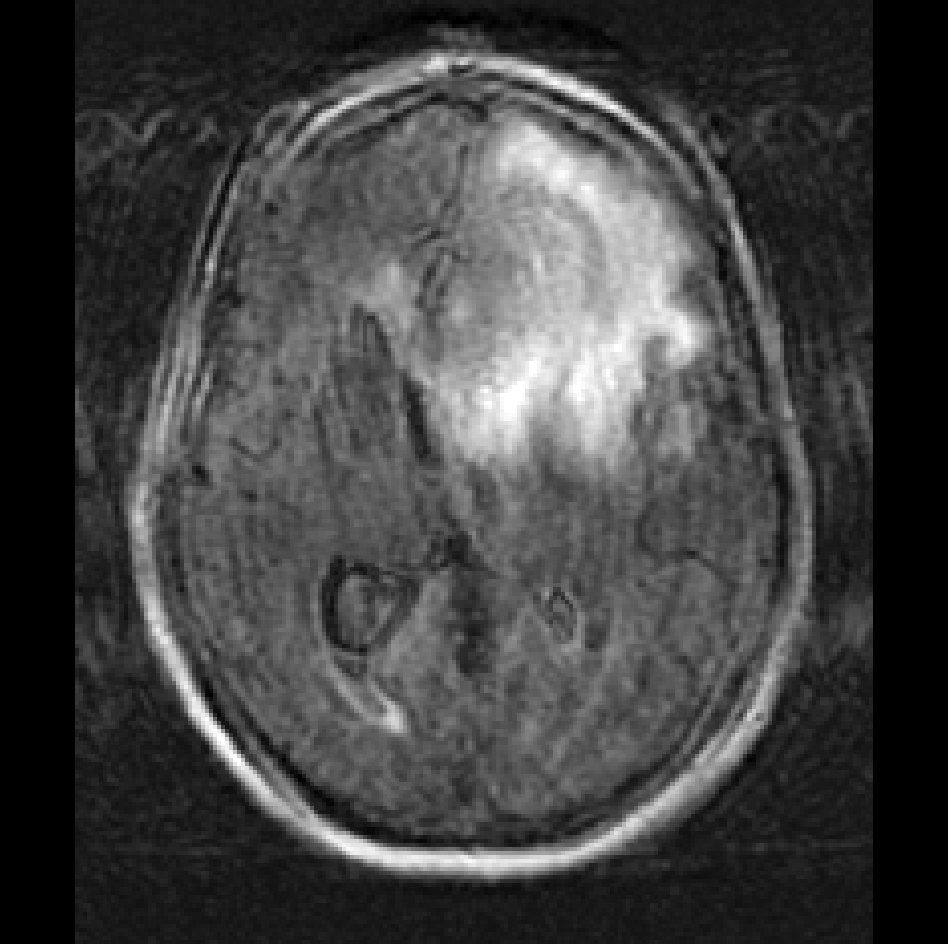}
        \hfill
        \includegraphics[width=0.43\textwidth, clip, trim=2.5cm 0cm 2.5cm 0cm]{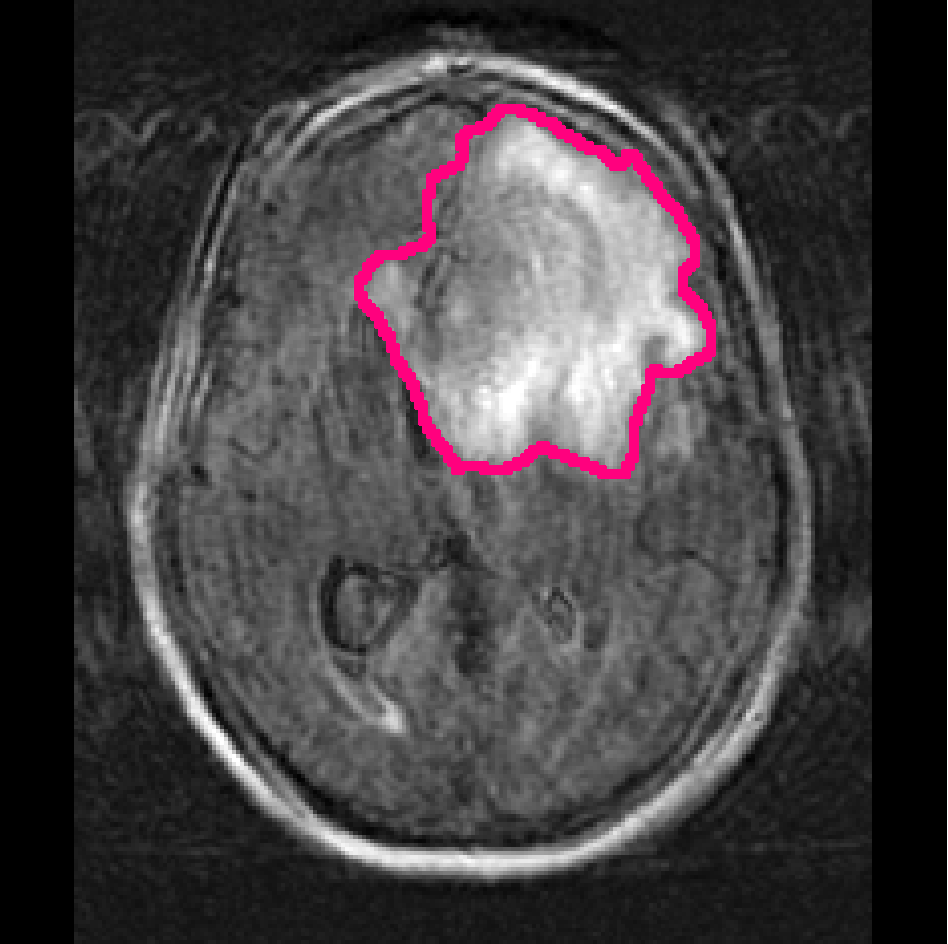}
        \hfill
    \end{subfigure}
    \caption{Example of a \acrshort{FLAIR} scan containing imaging artifacts, and the automatic segmentation (red overlay) made by our method.
    It was correctly predicted as an IDH wildtype, grade IV glioma. This is patient TCGA-06-5408 from the \acrshort{TCGAGBM} collection.}\label{fig:low_quality}
\end{figure}

By not excluding scans from our train set based on radiological characteristics, we were able to make our model robust to low scan quality, as can be seen in an example from the test set in \cref{fig:low_quality}.
Even though this example scan contained imaging artifacts, our method was able to properly segment the tumor (DICE score of 0.87), and correctly predict the tumor as an IDH wildtype, grade IV tumor.

\begin{figure}[htbp]
\centering
\begin{subfigure}[b]{\textwidth}
    \centering
    \hfill
    \begin{subfigure}[b]{0.24\textwidth}
    \includegraphics[width=\textwidth, clip, trim=2.5cm 0.5cm 2.5cm 0.5cm]{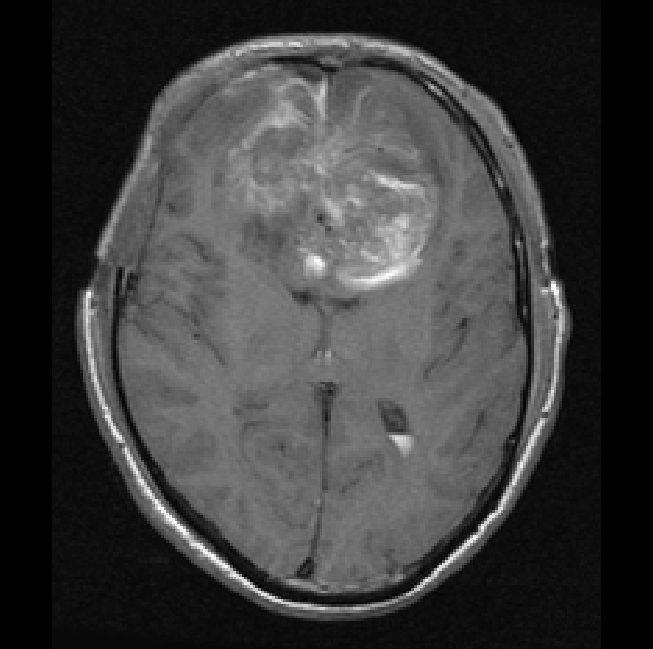}
    \caption*{Pre-contrast \acrshort{T1}}
    \end{subfigure}
    \hfill
    \begin{subfigure}[b]{0.24\textwidth}
    \includegraphics[width=\textwidth, clip, trim=2.5cm 0.5cm 2.5cm 0.5cm]{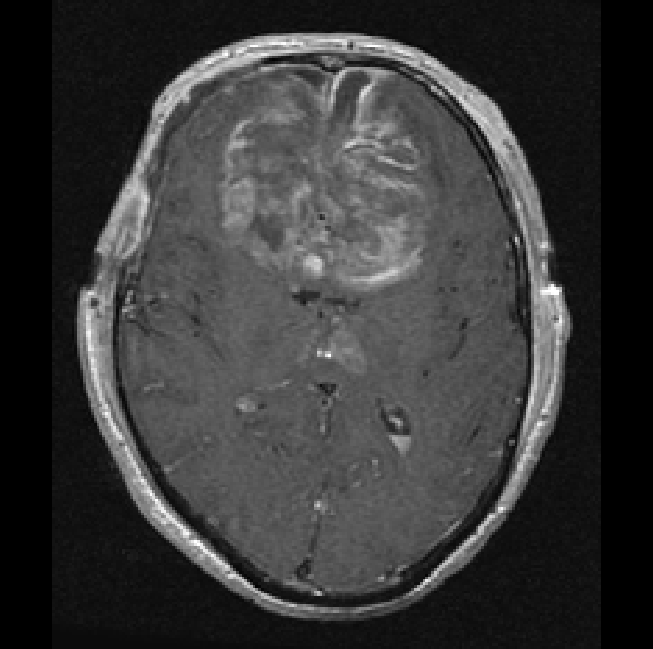}
    \caption*{Post-contrast \acrshort{T1}}
    \end{subfigure}
    \hfill
    \begin{subfigure}[b]{0.24\textwidth}
    \includegraphics[width=\textwidth, clip, trim=2.5cm 0.5cm 2.5cm 0.5cm]{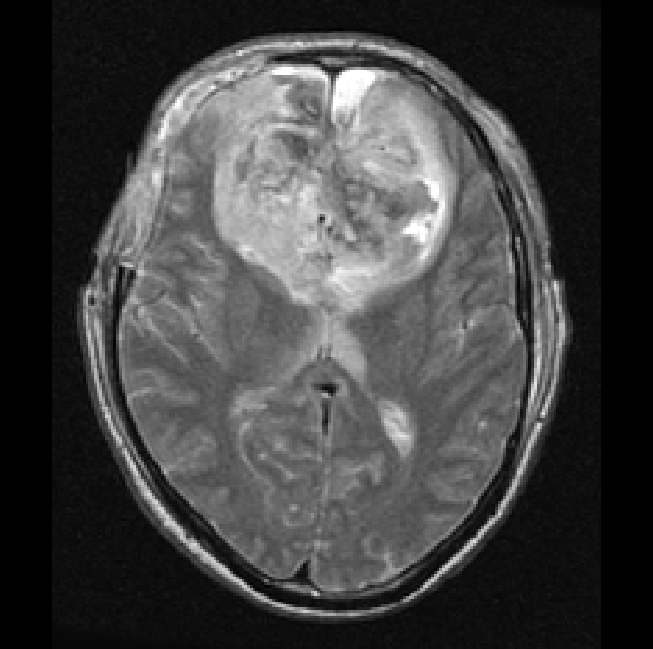}
    \caption*{\acrshort{T2}}
    \end{subfigure}
    \hfill
    \begin{subfigure}[b]{0.24\textwidth}
    \includegraphics[width=\textwidth, clip, trim=2.5cm 0.5cm 2.5cm 0.5cm]{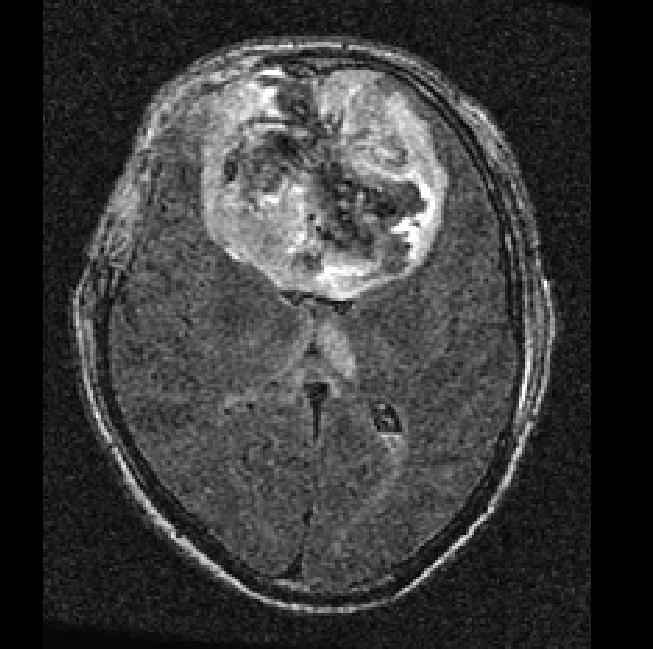}
    \caption*{\acrshort{FLAIR}}
    \end{subfigure}
    \caption{TCGA-DU-6410 from the \acrshort{TCGALGG} collection.
    The ground truth histopathological analysis indicated this glioma was grade II, \gls{IDH} mutated, 1p/19q co-deleted, but our method predicted it as a grade IV, \gls{IDH} wildtype.}\label{fig:wrong_prediction_LGG}
\end{subfigure}
\vskip\baselineskip
\begin{subfigure}[b]\textwidth
    \centering
    \hfill
    \begin{subfigure}[b]{0.24\textwidth}
    \includegraphics[width=\textwidth, clip, trim=2.5cm 0.5cm 2.5cm 0.5cm]{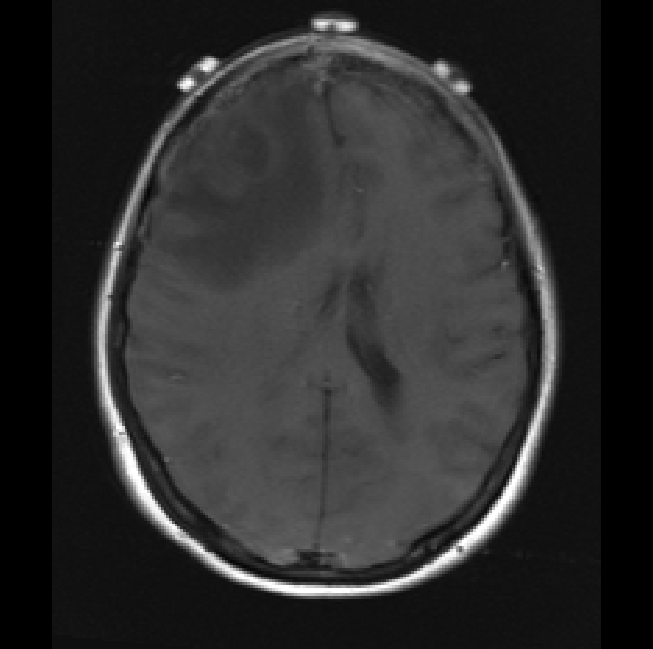}
    \caption*{Pre-contrast \acrshort{T1}}
    \end{subfigure}
    \hfill
    \begin{subfigure}[b]{0.24\textwidth}
    \includegraphics[width=\textwidth, clip, trim=2.5cm 0.5cm 2.5cm 0.5cm]{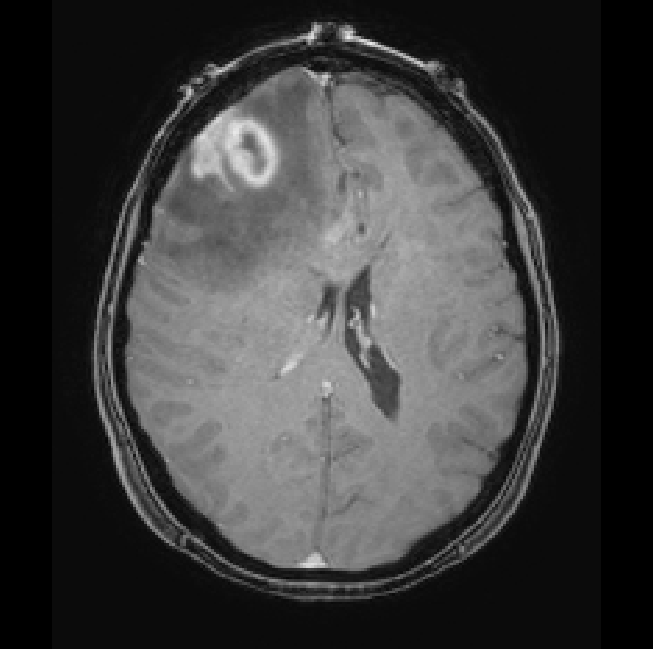}
    \caption*{Post-contrast \acrshort{T1}}
    \end{subfigure}
    \hfill
    \begin{subfigure}[b]{0.24\textwidth}
    \includegraphics[width=\textwidth, clip, trim=2.5cm 0.5cm 2.5cm 0.5cm]{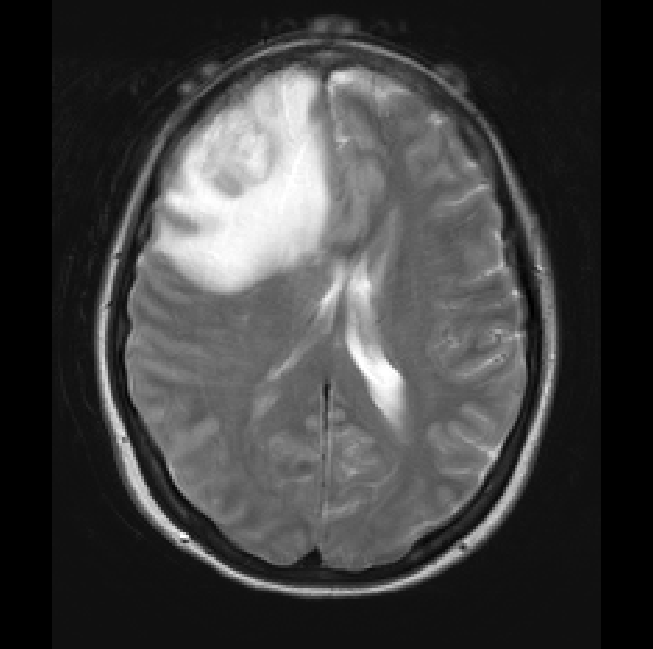}
    \caption*{\gls{T2}}
    \end{subfigure}
    \hfill
    \begin{subfigure}[b]{0.24\textwidth}
    \includegraphics[width=\textwidth, clip, trim=2.5cm 0.5cm 2.5cm 0.5cm]{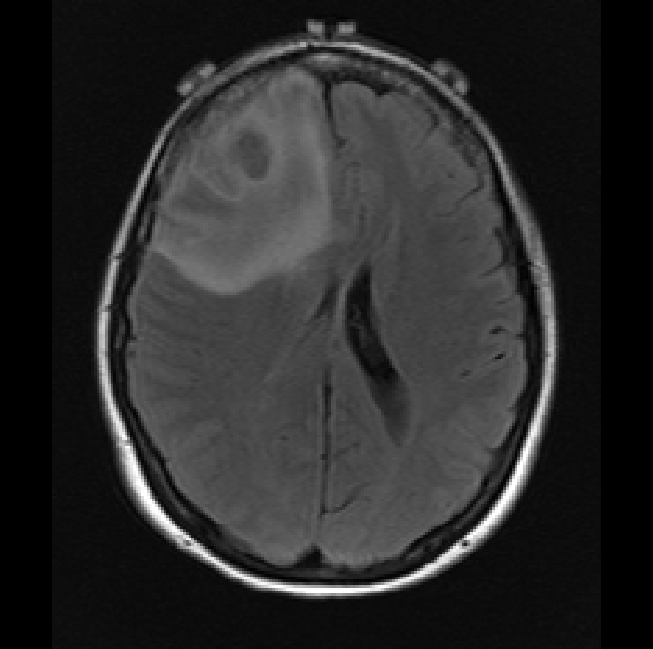}
    \caption*{\acrshort{FLAIR}}
    \end{subfigure}
    \caption{TCGA-76-7664 from the TCGA-HGG collection.
    Histopathologically this glioma was grade IV, \gls{IDH} wildtype, but our method predicted it as grade III, \gls{IDH} mutated.}\label{fig:wrong_prediction_HGG}
\end{subfigure}
\caption{Examples of scans that were incorrectly predicted by our method.}\label{fig:wrong_predictions}
\end{figure}

Finally, we considered two examples of scans that were incorrectly predicted by our method, see \cref{fig:wrong_predictions}.
These two examples were chosen because our network assigned high prediction scores to the wrong classes for these cases.
\cref{fig:wrong_prediction_LGG} shows an example of a  grade II, \gls{IDH} mutated, 1p/19q co-deleted glioma that was predicted as grade IV, \gls{IDH} wildtype by our method.
Our method's prediction was most likely caused by the hyperintensities in the post-contrast \gls{T1} scan being interpreted as contrast enhancement.
Since these hyperintensities are also present in the pre-contrast \gls{T1} scan they are most likely calcifications, and the radiological appearance of this tumor is indicative of an oligodendroglioma.
\cref{fig:wrong_prediction_HGG} shows an example of a grade IV, \gls{IDH} wildtype glioma that was predicted as a grade III, \gls{IDH} mutated glioma by our method.

\clearpage
\section{Discussion}

We have developed a method that can predict the IDH mutation status, 1p/19q co-deletion status, and grade of glioma, while simultaneously providing the tumor segmentation, based on pre-operative \gls{MRI} scans.
For the genetic and histological feature predictions, we achieved an AUC of 0.90 for the \gls{IDH} mutation status prediction, an \gls{AUC} of 0.85 for the 1p/19q co-deletion prediction, and an \gls{AUC} of 0.81 for the grade prediction, in the test set.

In an independent test set, which contained data from 13 different institutes, we demonstrated that our method predicts these features with good overall performance; we achieved an AUC of 0.90 for the \gls{IDH} mutation status prediction, an \gls{AUC} of 0.85 for the 1p/19q co-deletion prediction, and an \gls{AUC} of 0.81 for the grade prediction, and a mean whole tumor DICE score of 0.84.
This performance on unseen data that was only used during the final evaluation of the algorithm, and that was purposefully not used to guide any decisions regarding the method design, shows the true generalizability of our method.
Using the latest GPU capabilities we were able to train a large model, which uses the full 3D scan as input.
Furthermore, by using the largest, most diverse patient cohort to date we were able to make our method robust to the heterogeneity that is naturally present in clinical imaging data, such that it generalizes for broad application in clinical practice.

By using a multi-task network, our method could learn the context between different features.
For example, \gls{IDH} wildtype and 1p/19q co-deletion are mutually exclusive \cite{labussi2018idh}.
If two separate methods had been used, one to predict the IDH status and one to predict the 1p/19q co-deletion status, an IDH wildtype glioma might be predicted to be 1p/19q co-deleted, which does not stroke with the clinical reality.
Since our method learns both of these genetic features simultaneously, it correctly learned not to predict 1p/19q co-deletion in tumors that were IDH wildtype; there was only one patient in which our algorithm predicted a tumor to be both IDH wildtype and 1p/19q co-deleted.
Furthermore, by predicting the genetic and histological features individually, instead of only predicting the \gls{WHO} 2016 category, it is possible to adopt updated guidelines such as cIMPACT-NOW, future-proofing our method \cite{lous2020impactnow}.

Some previous studies also used multi-task networks to predict the genetic and histological features of glioma \cite{tang2020multi, xue2020radiomicsmulti, decuyper2020automated}.
\citeauthor{tang2020multi}~\cite{tang2020multi} used a multi-task network that predicts multiple genetic features, as well as the overall survival of glioblastoma.
Since their method only works for glioblastoma patients, the tumor grade must be known in advance, complicating the use of their method in the pre-operative setting when tumor grade is not yet known.
Furthermore, their method requires a tumor segmentation prior to application of their method, which is a time-consuming, expert task.
In a study by \citeauthor{xue2020radiomicsmulti}~\cite{xue2020radiomicsmulti}, a multi-task network was used, with a structure similar to the one proposed in this paper, to segment the tumor and predict the grade (\gls{LGG} or \gls{HGG}) and \gls{IDH} mutation status.
However, they do not predict the 1p/19q co-deletion status needed for the \gls{WHO} 2016 categorization.
Lastly, \citeauthor{decuyper2020automated}~\cite{decuyper2020automated} used a multi-task network that predicts the \gls{IDH} mutation and 1p/19q co-deletion status, and the tumor grade (\gls{LGG} or \gls{HGG}).
Their method requires a tumor segmentation as input, which they obtain from a U-Net that is applied earlier in their pipeline; thus, their method requires two networks instead of the single network we use in our method.
These differences aside, the most important limitation of each of these studies is the lack of an independent test set for evaluating their results.
It is now considered essential that an independent test set is used, to prevent an overly optimistic estimate of a method's performance \cite{gillies2016radiomics, rizzo2018radiomics, lohmann2020radiomics, yip2016applicationsradiomics}.
Thus, our study improves on this previous work by providing a single network that combines the different tasks, being trained on a more extensive and diverse dataset, not requiring a tumor segmentation as an input, providing all information needed for the WHO 2016 categorization, and, crucially, by being evaluated in an independent test set.

An important genetic feature that is not predicted by our method is the \gls{MGMT} methylation status.
Although the \gls{MGMT} methylation status is not part of the \gls{WHO} 2016 categorization, it is part of clinical management guidelines and is an important prognostic marker in glioblastoma \cite{louis20162016}.
In the initial stages of this study, we attempted to predict the \gls{MGMT} methylation status; however, the performance of this prediction was poor.
Furthermore, the methylation cutoff level, which is used to determine whether a tumor is \gls{MGMT} methylated, shows a wide variety between institutes, leading to inconsistent results  \cite{malstrom2019MGMT}.
We therefore opted not to include the \gls{MGMT} prediction at all, rather than to provide a poor prediction of an unsharply defined parameter.
Although some methods attempted to predict the \gls{MGMT} status, with varying degrees of success, there is still an ongoing discussion on the validity of MR imaging features of the \gls{MGMT} status \cite{tang2020multi, sasaki2019mgmt, gupta2013dwimgmt, carrillo2012mgmt, mikkelsen2020MGMT}.

Our method shows good overall performance, but there are noticeable performance differences between tumor categories.
For example, when our method predicts a tumor as an IDH wildtype glioblastoma, it is correct almost all of the time.
On the other hand, it has some difficulty differentiating IDH mutated, 1p/19q co-deleted low-grade glioma from other low-grade glioma.
The sensitivity for the prediction of grade III glioma was low, which might be caused by the lack of a central pathology review.
Because of this, there were differences in molecular testing and histological analysis, and it is known that distinguishing between grade II and grade III has a poor observer reliability \cite{vandenbent2010interobserver}.
Although our method can be relevant for certain subgroups, our method's performance still needs to be improved to ensure relevancy for the full patient population.

In future work, we aim to increase the performance of our method by including \gls{PWI} and \gls{DWI} since there has been an increasing amount of evidence that these physiological imaging modalities contain additional information that correlates with the tumor's genetic status and aggressiveness \cite{park2020radiomicspwi, kim2020diffusion}.
They were not included in this study since \gls{PWI} and, to a lesser extent, \gls{DWI} are not as ingrained in the clinical imaging routine as the structural scans used in this work \cite{thust2018gliomaimaging, fresychlag2018imaging}.
Thus, including these modalities would limit our method's clinical applicability and substantially reduce the number of patients in the train and test set.
However, \gls{PWI} and \gls{DWI} are increasingly becoming more commonplace, which will allow including these in future research and which might improve performance.

In conclusion, we have developed a non-invasive method that can predict the IDH mutation status, 1p/19q co-deletion status, and grade of glioma, while at the same time segmenting the tumor, based on pre-operative \gls{MRI} scans with high overall performance.
Although the performance of our method might need to be improved before it will find widespread clinical acceptance, we believe that this research is an important step forward in the field of radiomics.
Predicting multiple clinical features simultaneously steps away from the conventional single-task methods and is more in line with the clinical practice where multiple clinical features are considered simultaneously and may even be related.
Furthermore, by not limiting the patient population used to develop our method to a selection based on clinical or radiological characteristics, we alleviate the need for a priori (expert) knowledge, which may not always be available.
Although steps still have to be taken before radiomics will find its way into the clinic, especially in terms of performance, our work provides a crucial step forward by resolving some of the hurdles of clinical implementation now, and paving the way for a full transition in the future.

\section{Methods}

\subsection{Patient population}

The train set was collected from four in-house datasets and five publicly available datasets.
In-house datasets were collected from four different institutes: \gls{EMC}, \gls{HMC}, \gls{AMC} \cite{visser2019segmentation}, and \gls{UMCU}.
Four of the five public datasets were collected from \gls{TCIA} \cite{clark2013cancer}:  the \gls{REMBRANDT} collection \cite{scarpace2015radiology}, the \gls{CPTAC} collection \cite{cptac2018radiology}, the \gls{IVY} collection \cite{nameeta2016radiology, puchalski2018ivygap}, and the \gls{BTP} collection \cite{schmainda2018radiology}.
The fifth dataset was the 2019 \gls{BRATS} challenge dataset \cite{menze2015brats, bakas2017brats, bakas2018brats}, from which we excluded the patients that were also available in the \gls{TCGALGG} and \gls{TCGAGBM} collections \cite{pedano2016radiology, scarpace2016radiology}.

For the internal datasets from the \gls{EMC} and the \gls{HMC}, manual segmentations were available, which were made by four different clinical experts.
For patients where segmentations from more than one observer were available, we randomly picked one of the segmentations to use in the train set.
The segmentations from the \gls{AMC} data were made by a single observer of the study by \citeauthor{visser2019segmentation}~\cite{visser2019segmentation}.
From the public datasets, only the \gls{BRATS} dataset and the \gls{BTP} dataset provided manual segmentations.
Segmentations of the \gls{BRATS} dataset, as provided in the 2019 training and validation set were used.
For the \gls{BTP} dataset, the segmentations as provided in the \gls{TCIA} data collection were used.

Patients were included if pre-operative pre- and post-contrast \gls{T1}, \gls{T2}, and \gls{FLAIR} scans were available; no further inclusion criteria were set.
For example, patients were not excluded based on the radiological characteristics of the scan, such as low imaging quality or imaging artifacts, or the glioma's clinical characteristics such as the grade.
If multiple scans of the same contrast type were available in a single scan session (e.g., multiple \gls{T2} scans), the scan upon which the segmentation was made was selected.
If no segmentation was available, or the segmentation was not made based on that scan contrast, the scan with the highest axial resolution was used, where a 3D acquisition was preferred over a 2D acquisition.

For the in-house data, genetic and histological data were available for the \gls{EMC}, \gls{HMC}, and \gls{UMCU} dataset, which were obtained from analysis of tumor tissue after biopsy or resection.
Genetic and histological data of the public datasets were also available for the \gls{REMBRANDT}, \gls{CPTAC}, and \gls{IVY} collections.
Data for the \gls{REMBRANDT} and  \gls{CPTAC} collections was collected from the clinical data available at the \gls{TCIA} \cite{cptac2018radiology,scarpace2015radiology}.
For the \gls{IVY} collection, the genetic and histological data were obtained from the Swedish Institute at \url{https://ivygap.swedish.org/home}.

As a test set we used the \gls{TCGALGG} and \gls{TCGAGBM} collections from the \gls{TCIA} \cite{pedano2016radiology, scarpace2016radiology}.
Genetic and histological labels were obtained from the clinical data available at the \gls{TCIA}.
Segmentations were used as available from the \gls{TCIA}, based on the 2018 \gls{BRATS} challenge \cite{bakas2017brats, bakas2017segmentationLGG, bakas2017segmentationHGG}.
The inclusion criteria for the patients included in the \gls{BRATS} challenge were the same as our inclusion criteria: the presence of a pre-operative pre- and post-contrast \gls{T1}, \gls{T2}, and \gls{FLAIR} scan.
Thus, patients from the \gls{TCGALGG} and \gls{TCGAGBM} were included  if a segmentation from the \gls{BRATS} challenge was available.
However, for three patients, we found that although they did have manual segmentations, they did not meet our inclusion requirements: TCGA-08-0509 and TCGA-08-0510 from \gls{TCGAGBM} because they did not have a pre-contrast \gls{T1} scan and TCGA-FG-7634 from \gls{TCGALGG} because there was no post-contrast \gls{T1} scan.

\subsection{Automatic segmentation in the train set}

To present our method with a large diversity in scans, we wanted to include as many patients in the train set as possible from the different datasets.
Therefore, we performed automatic segmentation in patients that did not have manual segmentations.
To this end, we used an initial version of our network (presented in \cref{sec:model}), without the additional layers that were needed for the prediction of the genetic and histological features.
This network was initially trained using all patients in the train set for whom a manual segmentation was available, and this trained network was then applied to all patients for which a manual segmentation was not available.
The resulting automatic segmentations were inspected, and if their quality was acceptable, they were added to the train set.
The network was then trained again, using this increased dataset, and was applied to scans that did not yet have a segmentation of acceptable quality.
This process was repeated until an acceptable segmentation was available for all patients, which constituted our final, complete train set.

\subsection{Pre-processing}
For all datasets, except for the \gls{BRATS} dataset for which the scans were already provided in \acrshort{NIFTI} format, the scans were converted from DICOM format to \acrshort{NIFTI} format using dcm2niix version v1.0.20190410 \cite{li2016dcmniix}.
We then registered all scans to the MNI152 \gls{T1} and \gls{T2} atlases, version ICBM 2009a, which had a resolution of 1x1x1 mm$^3$ and a size of 197x233x189 voxels \cite{fonov2011unbiased, fonov2009unbiased}.
The scans were affinely registered using Elastix 5.0 \cite{klein2010elastix, shamonin2014fast}.
The pre- and post-contrast \gls{T1} scans were registered to the \gls{T1} atlas; the \gls{T2} and \gls{FLAIR} scans were registered to the \gls{T2} atlas.
When a manual segmentation was available for patients from the in-house datasets, the registration parameters that resulted from registering the scan used during the segmentation were used to transform the segmentation to the atlas.
In the case of the public datasets, we used the registration parameters of the \gls{FLAIR} scans to transform the segmentations.

After the registration, all scans were N4 bias field corrected using SimpleITK version 1.2.4 \cite{lowekamp2013simpleitk}.
A brain mask was made for the atlas using HD-BET, both for the \gls{T1} atlas and the \gls{T2} atlas \cite{isensee2019hdbet}.
This brain mask was used to skull strip all registered scans and crop them to a bounding box around the brain mask, reducing the amount of background present in the scans, resulting in a scan size of 152x182x145 voxels.
Subsequently, the scans were normalized such that for each scan, the average image intensity was 0, and the standard deviation of the image intensity was 1 within the brain mask.
Finally, the background outside the brain mask was set to the minimum intensity value within the brain mask.

Since the segmentation could sometimes be rugged at the edges after registration, especially when the segmentations were initially made on low-resolution scans, we smoothed the segmentation using a 3x3x3 median filter (this was only done in the train set).
For segmentations that contained more than one label, e.g., when the tumor necrosis and enhancement were separately segmented, all labels were collapsed into a single label to obtain a single segmentation of the whole tumor.
The genetic and histological labels and the segmentations of each patient were one-hot encoded.
The four scans, ground truth labels, and segmentation of each patient were then used as the input to the network.

\subsection{Model}\label{sec:model}

\begin{figure}
\includegraphics[width=\textwidth]{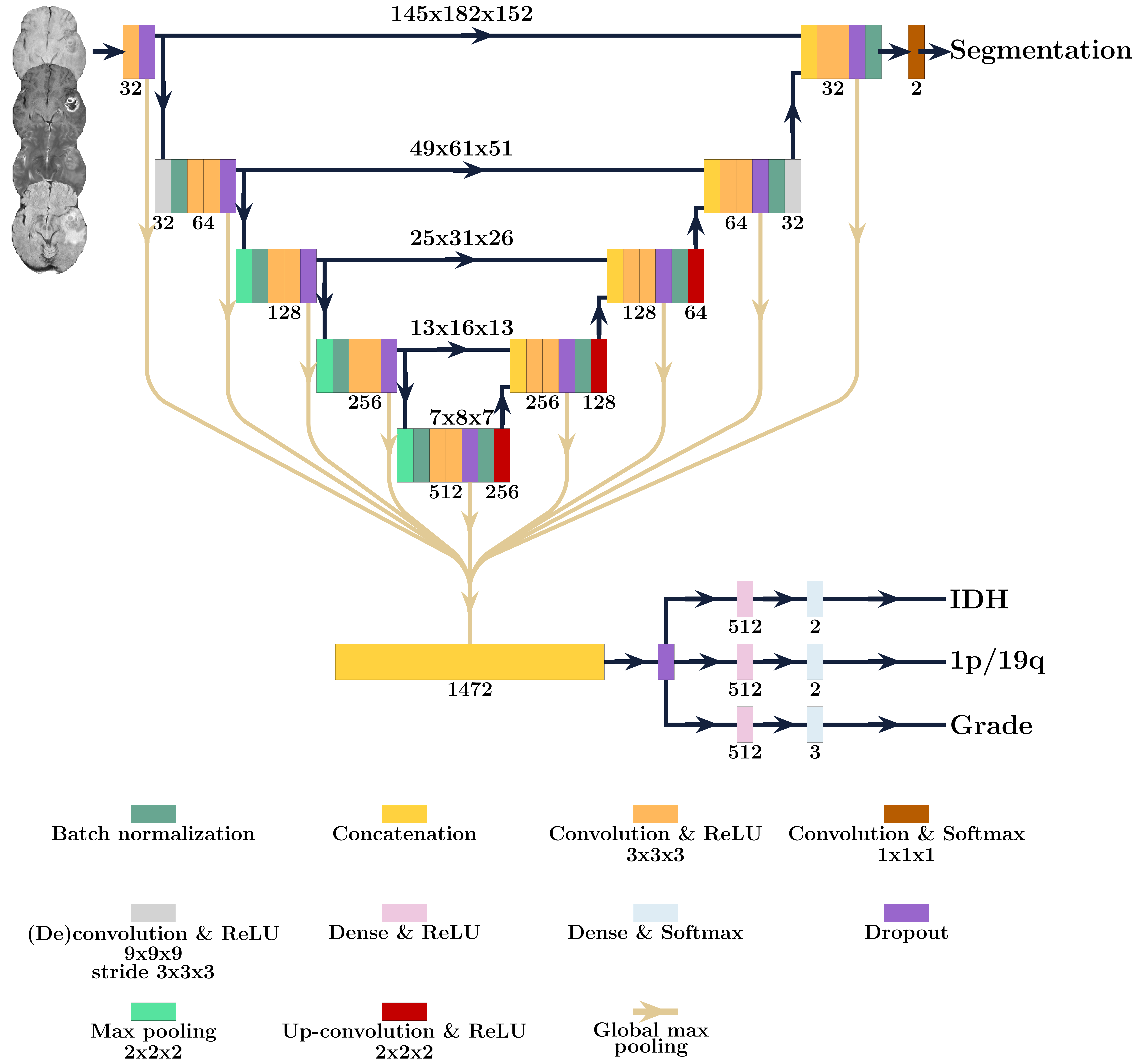}
\caption{Overview of the PrognosAIs Structure-Net (PS-Net) architecture used for our model. The numbers below the different layers indicate the number of filters, dense units or features at that layer.
We have also indicated the feature map size at the different depths of the network.}\label{fig:psnet_architecture}
\end{figure}

We based the architecture of our model on the U-Net architecture, with some adaptations made to allow for a full 3D input and the auxiliary tasks \cite{ronneberger2015unet}.
Our network architecture, which we have named PrognosAIs Structure-Net, or PS-Net for short, can be seen in \cref{fig:psnet_architecture}.

To use the full 3D scan as an input to the network, we replaced the first pooling layer that is usually present in the U-Net with a strided convolution, with a kernel size of 9x9x9 and a stride of 3x3x3.
In the upsampling branch of the network, the last up-convolution is replaced by a deconvolution, with the same kernel size and stride.

At each depth of the network, we have added global max-pooling layers directly after the dropout layer, to obtain imaging features that can be used to predict the genetic and histological features.
We chose global pooling layers as they do not introduce any additional parameters that need to be trained, thus keeping the memory required by our model manageable.
The features from the different depths of the network were concatenated and fed into three different dense layers, one for each of the genetic and histological outputs.

$l2$ kernel regularization was used in all convolutional layers, except for the last convolutional layer used for the output of the segmentation.
In total this model contained 27,042,473 trainable an 2,944 non-trainable parameters.

\subsection{Model training}

Training of the model was done on eight NVidia RTX2080Ti's with 11GB of memory, using TensorFlow 2.2.0 \cite{tensorflow2015whitepaper}.
To be able to use the full 3D scan as input to the network, without running into memory issues, we had to optimize the memory efficiency of the model training.
Most importantly, we used mixed-precision training, which means that most of the variables of the model (such as the weights) were stored in float16, which requires half the memory of float32, which is typically used to store these variables \cite{das2018mixed}.
Only the last softmax activation layers of each output were stored as float32.
We also stored our pre-processed scans as float16 to further reduce memory usage.

However, even with these settings, we could not use a batch size larger than 1.
It is known that a larger batch size is preferable, as it increases the stability of the gradient updates and allows for a better estimation of the normalization parameters in batch normalization layers \cite{smith2018batch}.
Therefore, we distributed the training over the eight GPUs, using the NCCL AllReduce algorithm, which combines the gradients calculated on each GPU before calculating the update to the model parameters \cite{woolleynccl}.
We also used synchronized batch normalization layers, which synchronize the updates of their parameters over the distributed models.
In this way, our model had a virtual batch size of eight for the gradient updates and the batch normalization layers parameters.

To provide more samples to the algorithm and prevent potential overtraining, we applied four types of data augmentation during training: cropping, rotation, brightness shifts, and contrast shifts.
Each augmentation was applied with a certain augmentation probability, which determined the probability of that augmentation type being applied to a specific sample.
When an image was cropped, a random number of voxels between 0 and 20 was cropped from each dimension, and filled with zeros.
For the random rotation, an angle between $-30^{\circ}$ and $30^{\circ}$ degrees was selected from a uniform distribution for each dimension.
The brightness shift was applied with a delta uniformly drawn between 0 and 0.2, and the contrast shift factor was randomly drawn between 0.85 and 1.15.
We also introduced an augmentation factor, which determines how often each sample was parsed as an input sample during a single epoch, where each time it could be augmented differently.

For the IDH, 1p/19q, and grade output, we used a masked categorical cross-entropy loss, and for the segmentation we used a DICE loss, see \cref{app:losses} for details.
We used AdamW as an optimizer, which has shown improved generalization performance over Adam by introducing the weight decay parameter as a separate parameter from the learning rate \cite{loshchilov2017decoupled}.
The learning rate was automatically reduced by a factor of 0.25 if the loss did not improve during the last five epochs, with a minimum learning rate of $1\cdot10^{-11}$.
The model could train for a maximum of 150 epochs, and training was stopped early if the average loss over the last five epochs did not improve.
Once the model was finished training, the weights from the epoch with the lowest loss were restored.

\subsection{Hyperparameter tuning}\label{sec:parameter_selection}

Hyperparameters involved in the training of the model needed to be tuned to achieve the best performance.
We tuned a total of six hyper parameters: the $l2$-norm, the dropout rate, the augmentation factor, the augmentation probability, the optimizer's initial learning rate and the optimizer's weight decay.
A full overview of the trained parameters and the values tested for the different settings is presented in \cref{app:hyperparameter_tuning}.

To tune these hyperparameters, we split the train set into a hyperparameter training set (85\%/1282 patients of the full train data) and a hyperparameter validation set (15\%/226 patients of the full train data).
Models were trained for different hyperparameter settings via  an exhaustive search using the hyperparameter train set, and then evaluated on the hyperparameter validation set.
No data augmentation was applied to the hyperparameter validation to ensure that results between trained models were comparable.
The hyperparameters that led to the lowest overall loss in the hyperparameter validation set were chosen as the optimal hyperparameters.
We trained the final model using these optimal hyperparameters and the full train set.

\subsection{Post-processing}

The predictions of the network were post-processed to obtain the final predicted labels and segmentations for the samples.
Since softmax activations were used for the genetic and histological outputs, a prediction between 0 and 1 was outputted for each class, where the individual predictions summed to 1.
The final predicted label was then considered as the class with the highest prediction score.
For the prediction of \gls{LGG} (grade II/III) vs. \gls{HGG} (grade IV), the prediction scores of grade II and grade III were combined to obtain the prediction score for \gls{LGG}, the prediction score of grade IV was used as the prediction score for \gls{HGG}.
If a segmentation contained multiple unconnected components, we only retained the largest component to obtain a single whole tumor segmentation.

\subsection{Model evaluation}

The performance of the final trained model was evaluated on the independent test set, comparing the predicted labels with the ground truth labels.
For the genetic and histological features, we evaluated the AUC, the accuracy, the sensitivity, and the specificity using scikit-learn version 0.23.1, for details see \cref{app:metric_defs} \cite{pedregosa2011scikitlearn}.
We evaluated these metrics on the full test set and in subcategories relevant to the \gls{WHO} 2016 guidelines.
We evaluated the \gls{IDH} performance separately in the \gls{LGG} (grade II/III) and \gls{HGG} (grade IV) subgroups, the 1p/19q performance in \gls{LGG}, and we also evaluated the performance of distinguishing between \gls{LGG} and \gls{HGG} instead of predicting the individual grades.

To evaluate the performance of the segmentation, we calculated the DICE scores, Hausdorff distances, and volumetric similarity coefficient comparing the automatic segmentation of our method and the manual ground truth segmentations for all patients in the test set.
These metrics were calculated using the EvaluateSegmentation toolbox, version 2017.04.25 \cite{taha2015metrics}, for details see \cref{app:metric_defs}.

To prevent an overly optimistic estimation of our model's predictive value, we only evaluated our model on the test set once all hyperparameters were chosen, and the final model was trained.
In this way, the performance in the test set did not influence decisions made during the development of the model, preventing possible overfitting by fine-tuning to the test set.

To gain insight into the model, we made saliency maps that show which parts of the scan contribute the most to the prediction of the \gls{CNN} \cite{smilkov2017smoothgrad}.
Saliency maps were made using tf-keras-vis 0.5.2, changing the activation function of all output layers from softmax to linear activations, using SmoothGrad to reduce the noisiness of the saliency maps \cite{smilkov2017smoothgrad}.

Another way to gain insight into the network's behavior is to visualize the filter outputs of the convolutional layers, as they can give some idea as to what operations the network applies to the scans.
We visualized the filter outputs of the last convolutional layers in the downsample and upsample path at the first depth (at an image size of 49x61x51) of our network.
These filter outputs were visualized by passing a sample through the network and showing the convolutional layers' outputs, replacing the ReLU activation with linear activations.

\subsection{Data availability}
An overview of the patients included from the public datasets used in the training and testing of the algorithm, and their ground truth label is available in \cref{app:open_gt}.
The data from the public datasets are available in TCIA under DOIs: 10.7937/K9/TCIA.2015.588OZUZB, 10.7937/k9/tcia.2018.3rje41q1, 10.7937/K9/TCIA.2016.XLwaN6nL, and 10.7937/K9/TCIA.2018.15quzvnb.
Data from the \gls{BRATS} are available at \url{http://braintumorsegmentation.org/}.
Data from the in-house datasets are not publicly available due to participant privacy and consent.

\subsection{Code availability}
The code used in this paper is available on GitHub under an Apache 2 license at \url{https://github.com/Svdvoort/PrognosAIs_glioma}.
This code includes the full pipeline from registration of the patients to the final post-processing of the predictions.
The trained model is also available on GitHub, along with code to apply it to new patients.

\newpage
\appendix
\section*{Appendices}

\section{Confusion matrices}\label{app:confusion_matrix}

\cref{tab:conf_IDH,tab:conf_1p19q,tab:conf_grade} show the confusion matrices for the \gls{IDH}, 1p/19q, and grade predictions, and \cref{tab:conf_who} shows the confusion matrix for the WHO 2016 subtypes.

\cref{tab:conf_1p19q} shows that the algorithm mainly has difficulty recognizing  1p/19q co-deleted tumors, which are mostly predicted as 1p/19q intact.
\cref{tab:conf_grade} shows that most of the incorrectly predicted grade III tumors are predicted as grade IV tumors.

\cref{tab:conf_who} shows that our algorithm often incorrectly predicts IDH-wildtype astrocytoma as IDH-wildtype glioblastoma.
The latest cIMPACT-NOW guidelines propose a new categorization, in which IDH-wildtype astrocytoma that show either TERT promoter methylation, or EFGR gene amplification, or chromosome 7 gain/chromsome 10 loss are classified as IDH-wildtype glioblastoma \cite{lous2020impactnow}.
This new categorization is proposed since the survival of patients with those IDH-wildtype astrocytoma is similar to the survival of patients with IDH-wildtype glioblastoma \cite{lous2020impactnow}.
From the 13 IDH-wildtype astrocytoma that were wrongly predicted as IDH-wildtype glioblastoma, 12 would actually be categorized as IDH-wildtype glioblastoma under this new categorization.
Thus, although our method wrongly predicted the WHO 2016 subtype, it might actually have picked up on imaging features related to the aggressiveness of the tumor, which might lead to a better categorization.

{    
\begin{table}[htbp]
\caption{Confusion matrix of the IDH predictions.}\label{tab:conf_IDH}
\makegapedcells
\begin{tabular}{cc|cc}
\multicolumn{2}{c}{}
            &   \multicolumn{2}{c}{\textbf{Predicted}} \\
    &       &   Wildtype &   Mutated              \\
    \cline{2-4}
\multirow{2}{*}{\rotatebox[origin=c]{90}{\textbf{Actual}}}
    & Wildtype   & 120   & 9                 \\
    & Mutated    & 25    & 63                \\
    \cline{2-4}
    \end{tabular}
\end{table}
 }

 {    
\begin{table}[htbp]
    \caption{Confusion matrix of the 1p/19q predictions.}\label{tab:conf_1p19q}
\makegapedcells
\begin{tabular}{cc|cc}
\multicolumn{2}{c}{}
            &   \multicolumn{2}{c}{\textbf{Predicted}} \\
    &       &   Intact &   Co-deleted              \\
    \cline{2-4}
\multirow{2}{*}{\rotatebox[origin=c]{90}{\textbf{Actual}}}
    & Intact   & 197   & 10                \\
    & Co-deleted    & 16    & 10                \\
    \cline{2-4}
    \end{tabular}
\end{table}
 }

 {    
 \begin{table}[htbp]
    \caption{Confusion matrix of the grade predictions.}\label{tab:conf_grade}
 \makegapedcells
 \begin{tabular}{cc|ccc}
 \multicolumn{2}{c}{}
             &   \multicolumn{3}{c}{\textbf{Predicted}} \\
     &       &   Grade II &   Grade III & Grade IV              \\
     \cline{2-5}
 \multirow{3}{*}{\rotatebox[origin=c]{90}{\textbf{Actual}}}
     & Grade II   & 35   & 6 & 6                \\
     & Grade III    & 19 & 10 & 30                \\
     & Grade IV    & 2    & 5 & 125                \\
     \cline{2-5}
     \end{tabular}
 \end{table}
  }

  {    
  \begin{table}[htbp]
     \caption{Confusion matrix of the WHO 2016 predictions.
     The 'other' category indicates patients that were predicted as a non-existing WHO 2016 subtype, for example IDH wildtype, 1p/19q co-deleted tumors.
      Only one patient (TCGA-HT-A5RC) was predicted as a non-existing category. It was predicted as an IDH wildtype, 1p/19q co-deleted, grade IV tumor.}\label{tab:conf_who}
  \makegapedcells
  \hspace*{-3cm}
  \begin{tabular}{cc|cccccc}
  \multicolumn{2}{c}{}
              &   \multicolumn{6}{c}{\textbf{Predicted}} \\
      &       &   \rotatebox{-45}{Oligodendroglioma} & \rotatebox{-45}{\shortstack{IDH-mutated\\astrocytoma}} & \rotatebox{-45}{\shortstack{IDH-wildtype\\astrocytoma}}  &  \rotatebox{-45}{\shortstack{IDH-mutated\\glioblastoma}} & \rotatebox{-45}{\shortstack{IDH-wildtype\\glioblastoma}} \rotatebox{-45}{Other}              \\
      \cline{2-8}
  \multirow{5}{*}{\rotatebox[origin=c]{90}{\textbf{Actual}\hphantom{aaaa}}}
      & Oligodendroglioma    & 10    & 8 & 1 & 0 & 7 &0                \\[2ex]
      & \shortstack{IDH-mutated\\astrocytoma}    & 6 & 34 & 4 & 3 & 10 &0                \\[2ex]
      & \shortstack{IDH-wildtype\\astrocytoma}   & 1   & 2 & 3 & 2 & 13 & 1                \\[2ex]
      &\shortstack{IDH-mutated\\glioblastoma} & 0 & 1 & 0 & 0 & 3 & 0\\[2ex]
      & \shortstack{IDH-wildtype\\glioblastoma} & 0 & 3 &  3 & 1 & 96 & 0 \\[2ex]
      \cline{2-8}
      \end{tabular}
      {\small \raggedright Oligodendroglioma are IDH-mutated, 1p/19q co-deleted, grade II/III gioma.\\
      IDH-mutated astrocytoma are IDH-mutated, 1p/19q intact, grade II/III glioma.\\
      IDH-wildtype astrocytoma are IDH-wildtype, 1p/19q intact, grade II/III glioma.\\
      IDH-mutated glioblastoma are IDH-mutated, grade IV glioma.\\
      IDH-wildtype glioblastoma are IDH-wildtype, grade IV glioma.  \par}
  \end{table}
   }

\clearpage
\newpage

\section{Segmentation examples}\label{app:seg_examples_random}
\begin{figure}[htbp]
    \centering
    \begin{subfigure}[b]{\textwidth}
        \centering
        \hfill
        \begin{subfigure}[b]{0.215\textwidth}
        \caption*{Pre-contrast \acrshort{T1}}
        \includegraphics[width=\textwidth, clip, trim=2.5cm 0.5cm 2.5cm 0.5cm]{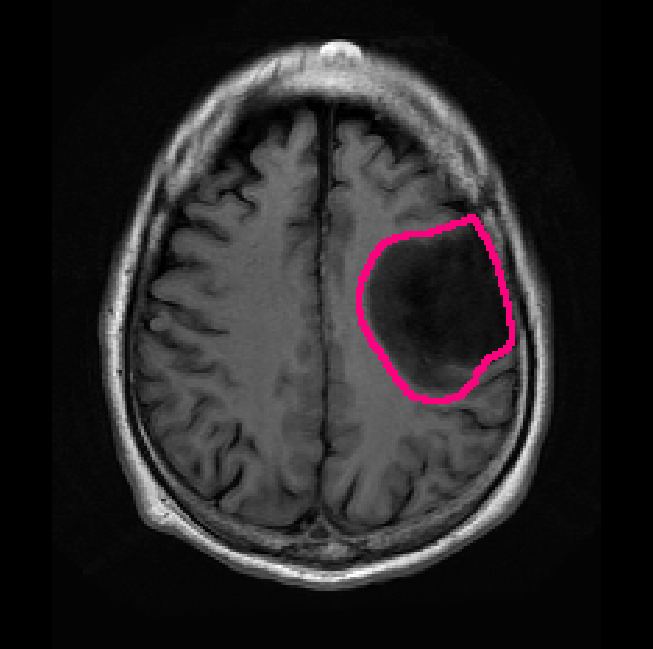}
        \end{subfigure}
        \hfill
        \begin{subfigure}[b]{0.215\textwidth}
        \caption*{Post-contrast \acrshort{T1}}
        \includegraphics[width=\textwidth, clip, trim=2.5cm 0.5cm 2.5cm 0.5cm]{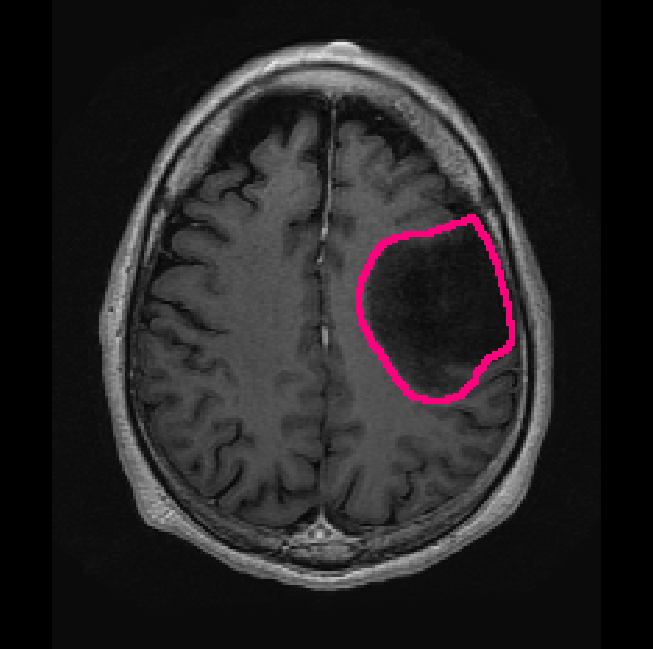}
        \end{subfigure}
        \hfill
        \begin{subfigure}[b]{0.215\textwidth}
        \caption*{\acrshort{T2}}
        \includegraphics[width=\textwidth, clip, trim=2.5cm 0.5cm 2.5cm 0.5cm]{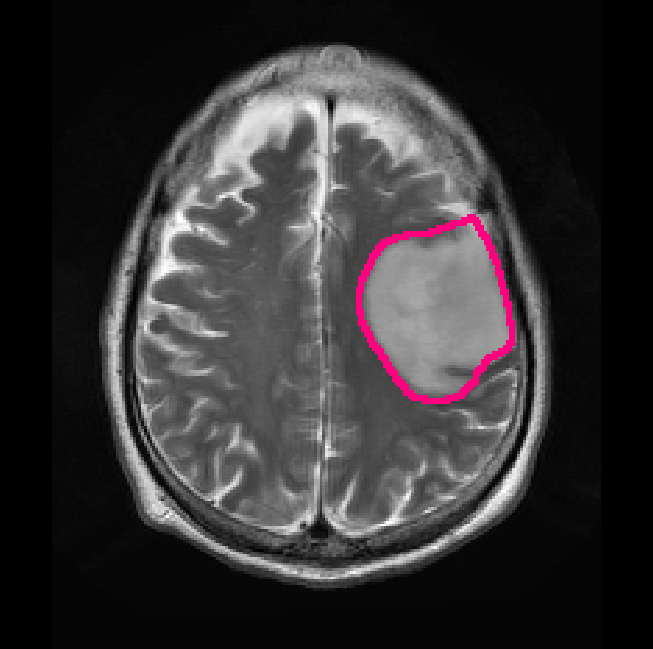}
        \end{subfigure}
        \hfill
        \begin{subfigure}[b]{0.215\textwidth}
        \caption*{\acrshort{FLAIR}}
        \includegraphics[width=\textwidth, clip, trim=2.5cm 0.5cm 2.5cm 0.5cm]{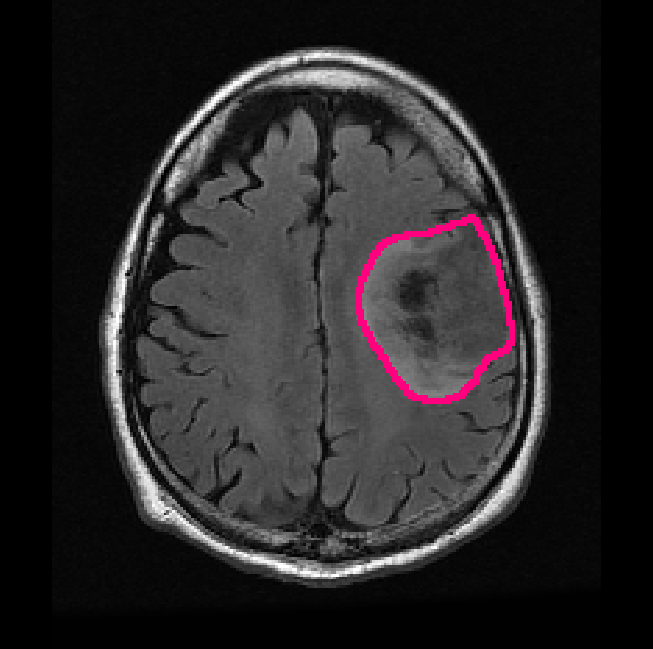}
        \end{subfigure}
        \caption{Patient TCGA-DU-7301 from the TCGA-LGG data collection.}
    \end{subfigure}
    \begin{subfigure}[b]\textwidth
        \centering
        \hfill
        \begin{subfigure}[b]{0.215\textwidth}
        \includegraphics[width=\textwidth, clip, trim=2.5cm 0.5cm 2.5cm 0.5cm]{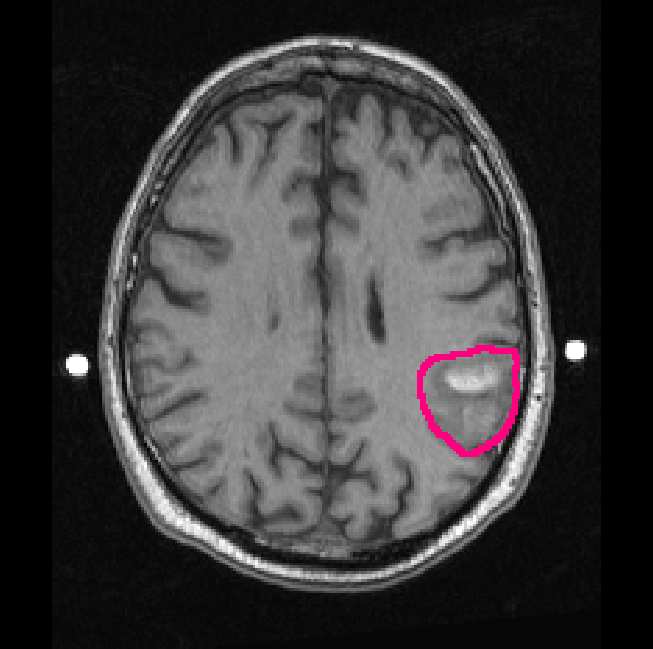}
        \end{subfigure}
        \hfill
        \begin{subfigure}[b]{0.215\textwidth}
        \includegraphics[width=\textwidth, clip, trim=2.5cm 0.5cm 2.5cm 0.5cm]{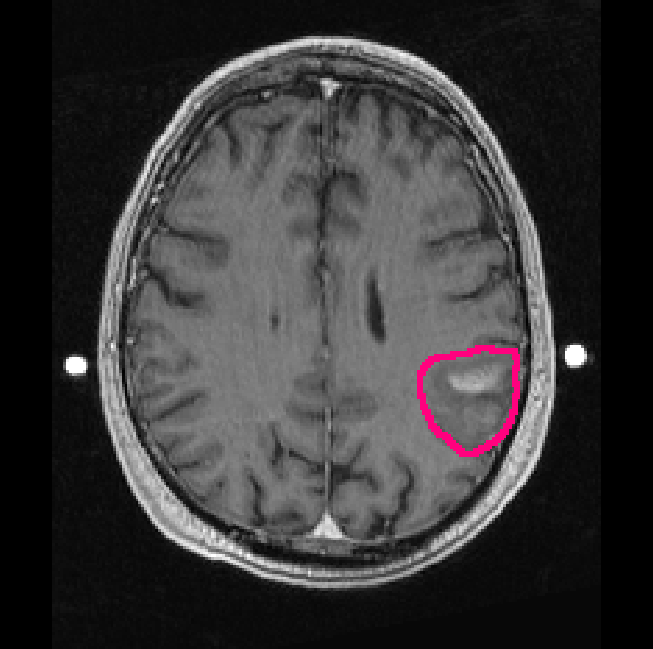}
        \end{subfigure}
        \hfill
        \begin{subfigure}[b]{0.215\textwidth}
        \includegraphics[width=\textwidth, clip, trim=2.5cm 0.5cm 2.5cm 0.5cm]{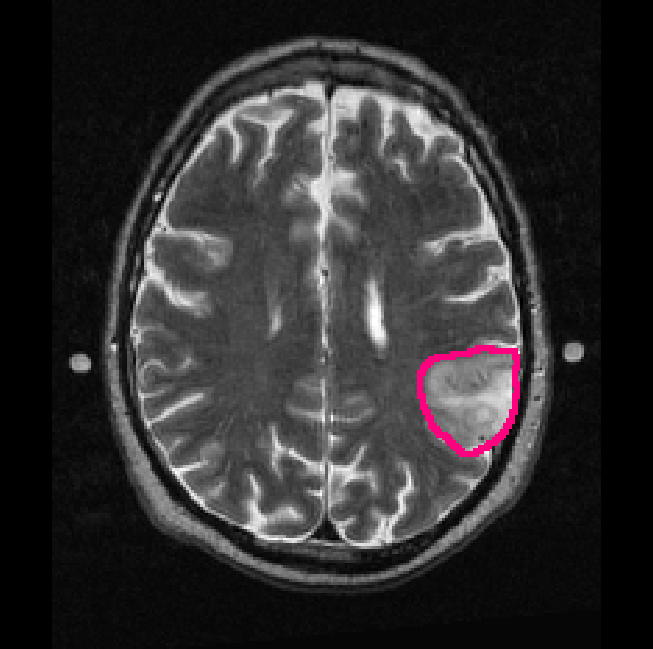}
        \end{subfigure}
        \hfill
        \begin{subfigure}[b]{0.215\textwidth}
        \includegraphics[width=\textwidth, clip, trim=2.5cm 0.5cm 2.5cm 0.5cm]{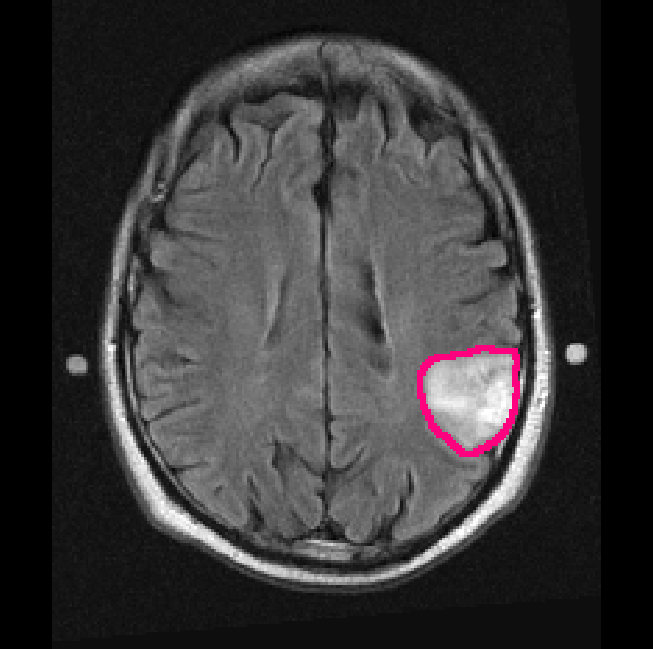}
        \end{subfigure}
        \caption{Patient TCGA-FG-5964 from the TCGA-LGG data collection.}
    \end{subfigure}
    \begin{subfigure}[b]\textwidth
        \centering
        \hfill
        \begin{subfigure}[b]{0.215\textwidth}
        \includegraphics[width=\textwidth, clip, trim=2.5cm 0.5cm 2.5cm 0.5cm]{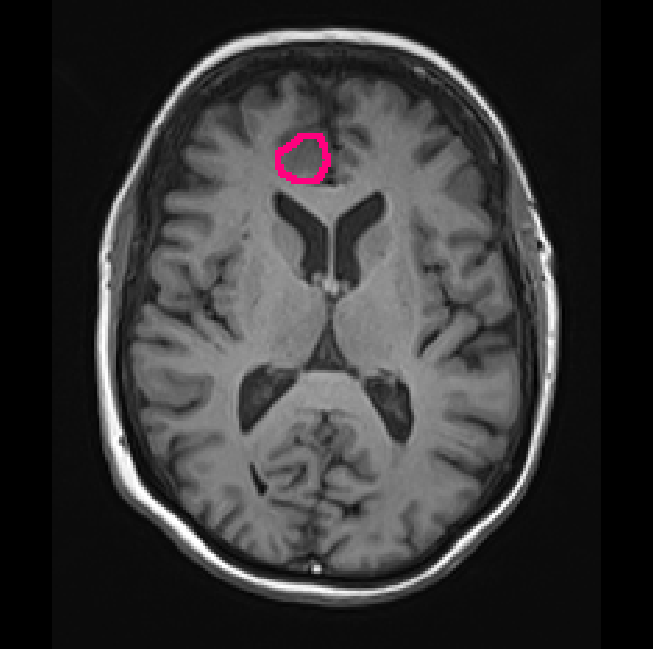}
        \end{subfigure}
        \hfill
        \begin{subfigure}[b]{0.215\textwidth}
        \includegraphics[width=\textwidth, clip, trim=2.5cm 0.5cm 2.5cm 0.5cm]{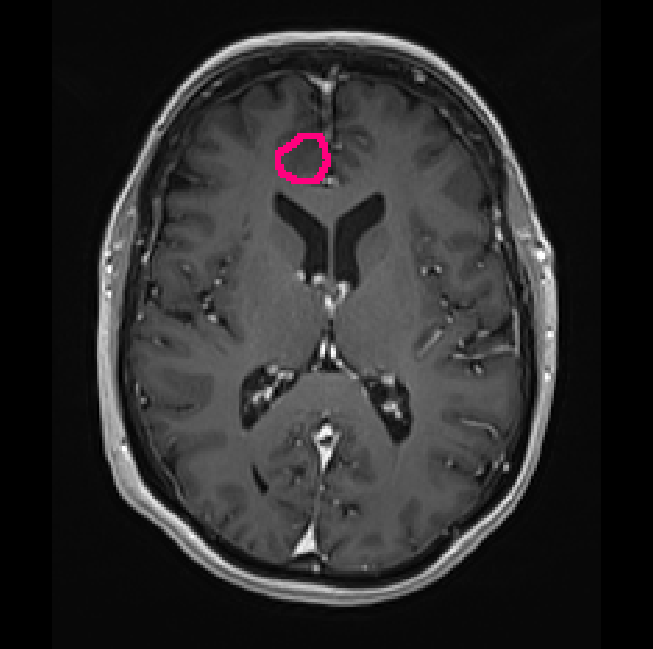}
        \end{subfigure}
        \hfill
        \begin{subfigure}[b]{0.215\textwidth}
        \includegraphics[width=\textwidth, clip, trim=2.5cm 0.5cm 2.5cm 0.5cm]{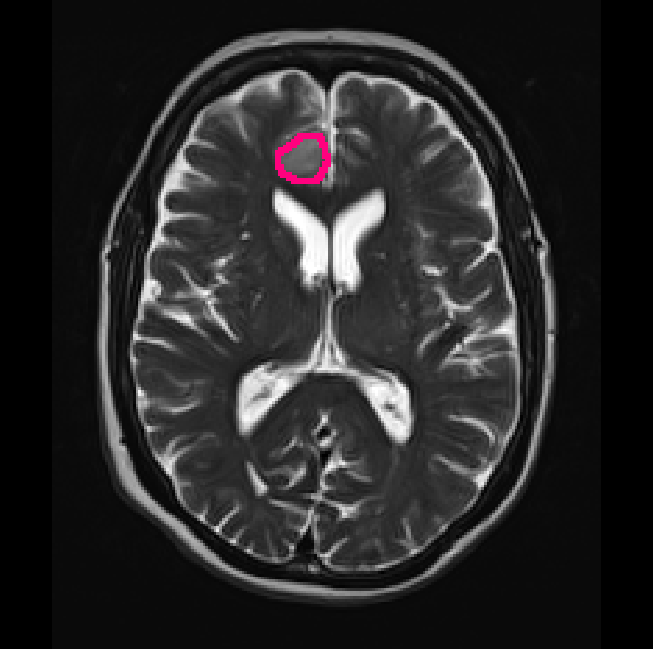}
        \end{subfigure}
        \hfill
        \begin{subfigure}[b]{0.215\textwidth}
        \includegraphics[width=\textwidth, clip, trim=2.5cm 0.5cm 2.5cm 0.5cm]{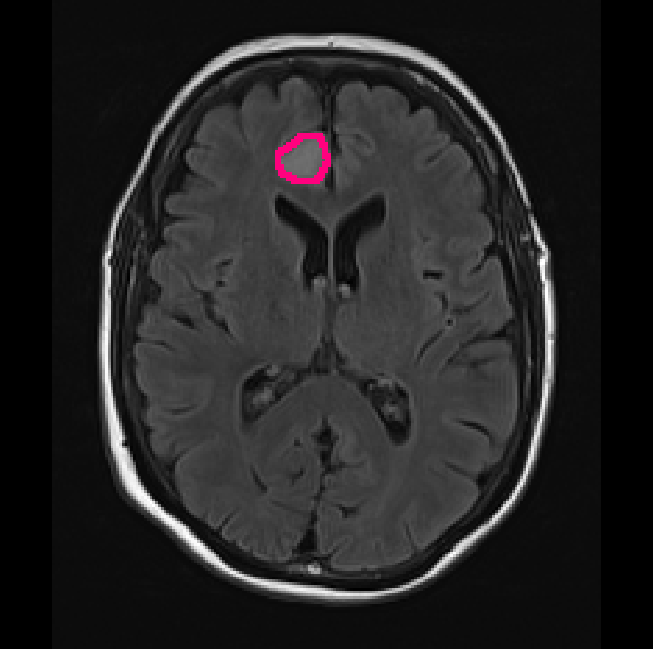}
        \end{subfigure}
        \caption{Patient TCGA-FG-A713 from the TCGA-LGG data collection.}
    \end{subfigure}
    \begin{subfigure}[b]\textwidth
        \centering
        \hfill
        \begin{subfigure}[b]{0.215\textwidth}
        \includegraphics[width=\textwidth, clip, trim=2.5cm 0.5cm 2.5cm 0.5cm]{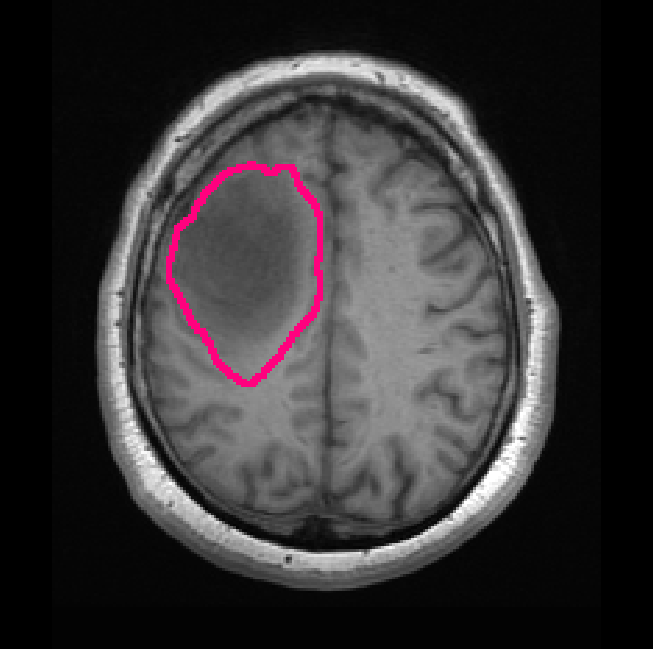}
        \end{subfigure}
        \hfill
        \begin{subfigure}[b]{0.215\textwidth}
        \includegraphics[width=\textwidth, clip, trim=2.5cm 0.5cm 2.5cm 0.5cm]{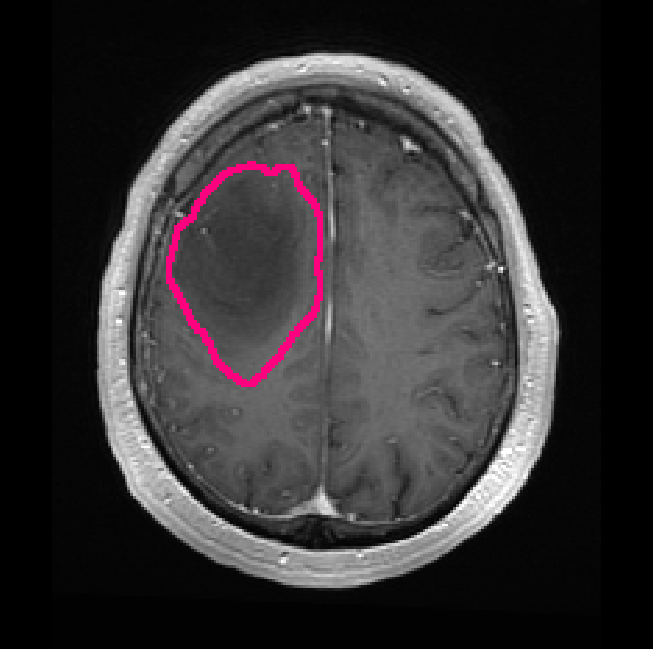}
        \end{subfigure}
        \hfill
        \begin{subfigure}[b]{0.215\textwidth}
        \includegraphics[width=\textwidth, clip, trim=2.5cm 0.5cm 2.5cm 0.5cm]{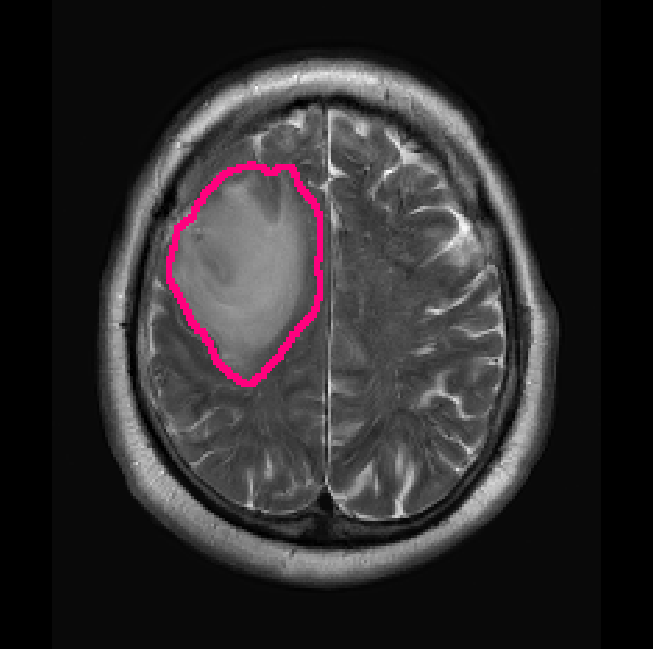}
        \end{subfigure}
        \hfill
        \begin{subfigure}[b]{0.215\textwidth}
        \includegraphics[width=\textwidth, clip, trim=2.5cm 0.5cm 2.5cm 0.5cm]{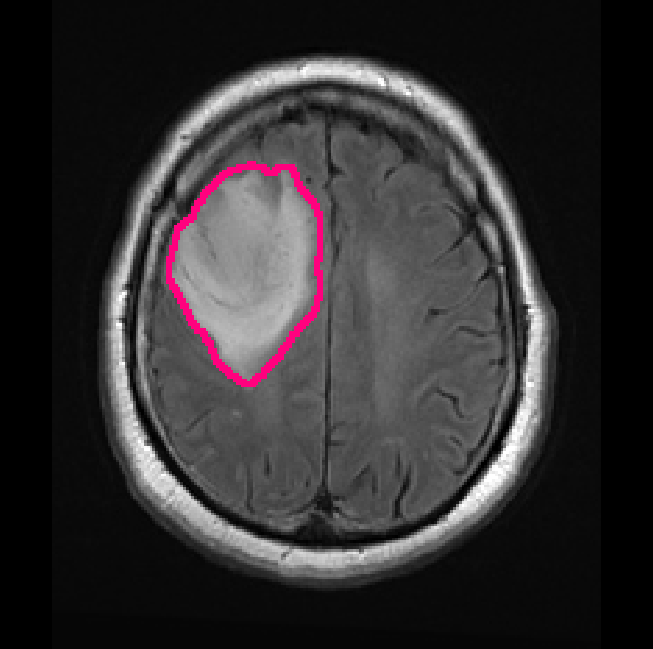}
        \end{subfigure}
        \caption{Patient TCGA-HT-7475 from the TCGA-LGG data collection.}
    \end{subfigure}
    \begin{subfigure}[b]\textwidth
        \centering
        \hfill
        \begin{subfigure}[b]{0.215\textwidth}
        \includegraphics[width=\textwidth, clip, trim=2.5cm 0.5cm 2.5cm 0.5cm]{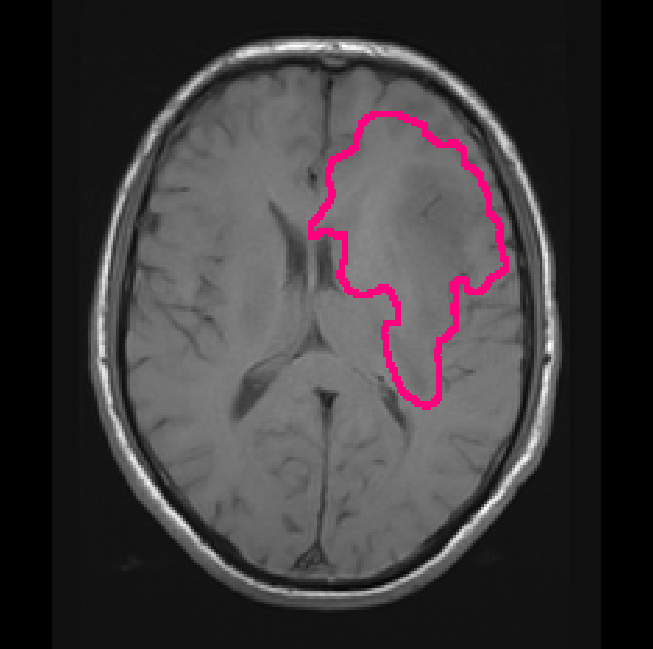}
        \end{subfigure}
        \hfill
        \begin{subfigure}[b]{0.215\textwidth}
        \includegraphics[width=\textwidth, clip, trim=2.5cm 0.5cm 2.5cm 0.5cm]{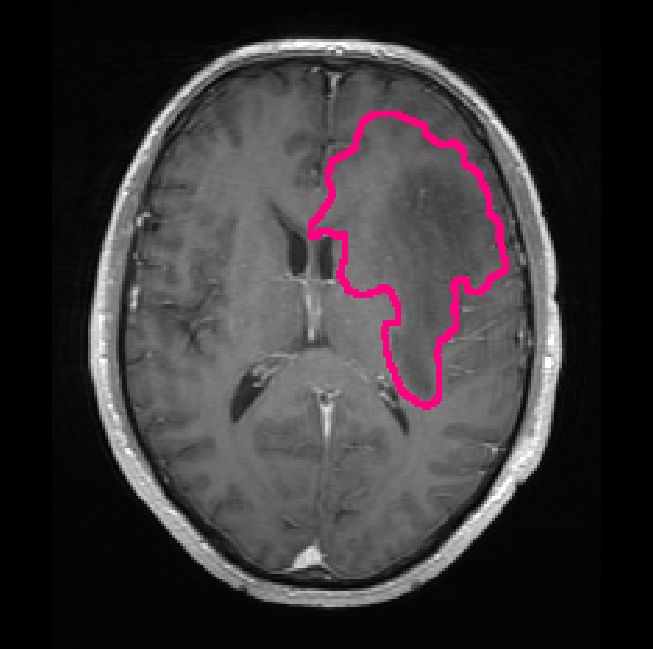}
        \end{subfigure}
        \hfill
        \begin{subfigure}[b]{0.215\textwidth}
        \includegraphics[width=\textwidth, clip, trim=2.5cm 0.5cm 2.5cm 0.5cm]{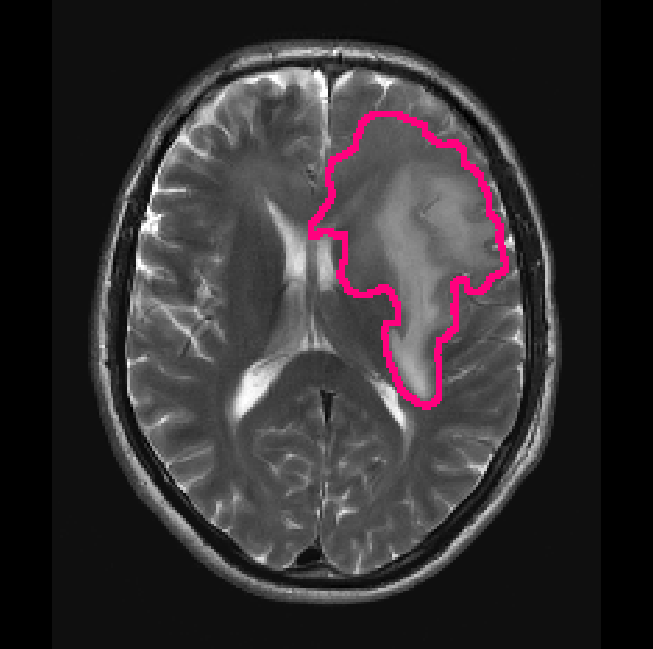}
        \end{subfigure}
        \hfill
        \begin{subfigure}[b]{0.215\textwidth}
        \includegraphics[width=\textwidth, clip, trim=2.5cm 0.5cm 2.5cm 0.5cm]{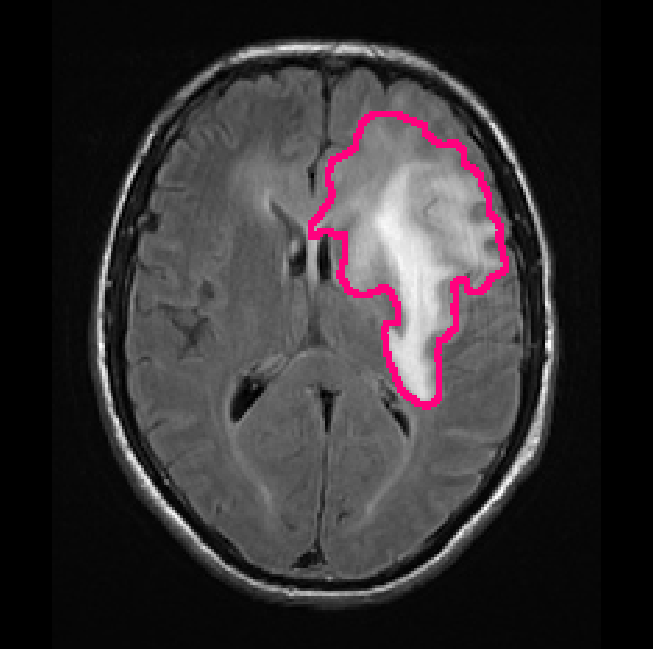}
        \end{subfigure}
        \caption{Patient TCGA-HT-8106 from the TCGA-LGG data collection.}
    \end{subfigure}
    \caption{Examples of scans and automatic segmentations of five patients that were randomly selected from the TCGA-LGG data collection.}\label{fig:seg_examples_LGG}
\end{figure}

\begin{figure}[htbp]
    \centering
    \begin{subfigure}[b]{\textwidth}
        \centering
        \hfill
        \begin{subfigure}[b]{0.215\textwidth}
        \caption*{Pre-contrast \acrshort{T1}}
        \includegraphics[width=\textwidth, clip, trim=2.5cm 0.5cm 2.5cm 0.5cm]{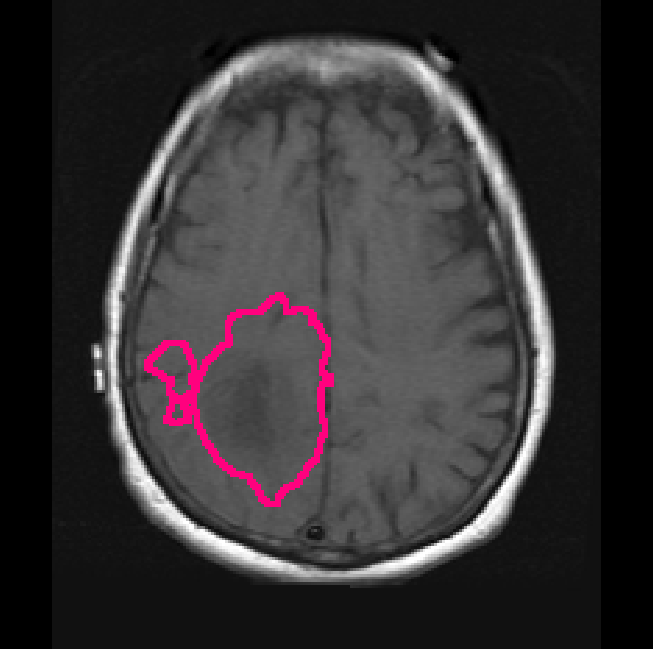}
        \end{subfigure}
        \hfill
        \begin{subfigure}[b]{0.215\textwidth}
        \caption*{Post-contrast \acrshort{T1}}
        \includegraphics[width=\textwidth, clip, trim=2.5cm 0.5cm 2.5cm 0.5cm]{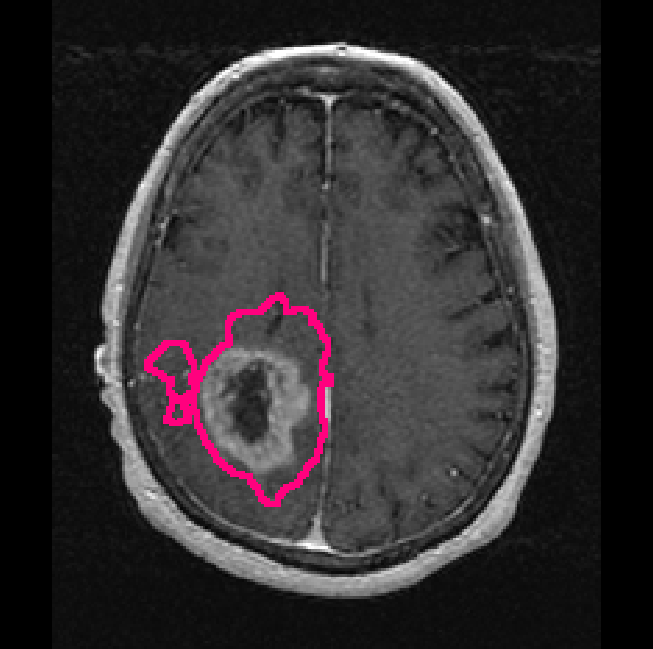}
        \end{subfigure}
        \hfill
        \begin{subfigure}[b]{0.215\textwidth}
        \caption*{\acrshort{T2}}
        \includegraphics[width=\textwidth, clip, trim=2.5cm 0.5cm 2.5cm 0.5cm]{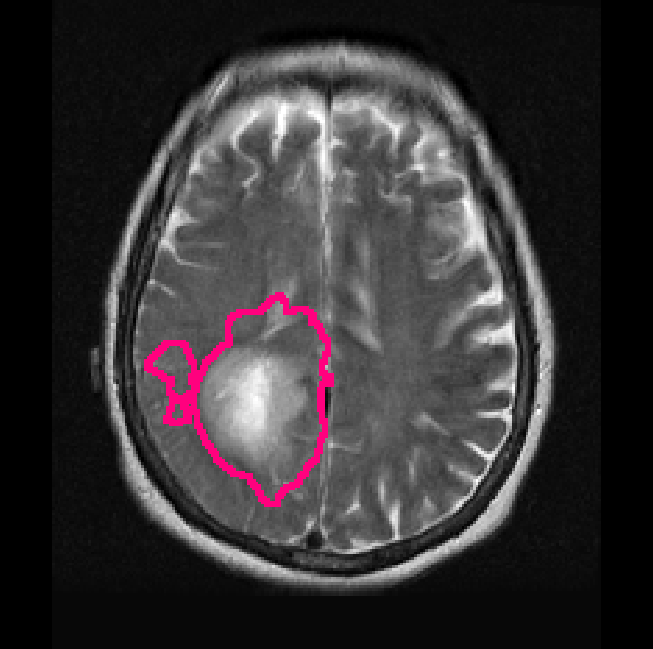}
        \end{subfigure}
        \hfill
        \begin{subfigure}[b]{0.215\textwidth}
        \caption*{\acrshort{FLAIR}}
        \includegraphics[width=\textwidth, clip, trim=2.5cm 0.5cm 2.5cm 0.5cm]{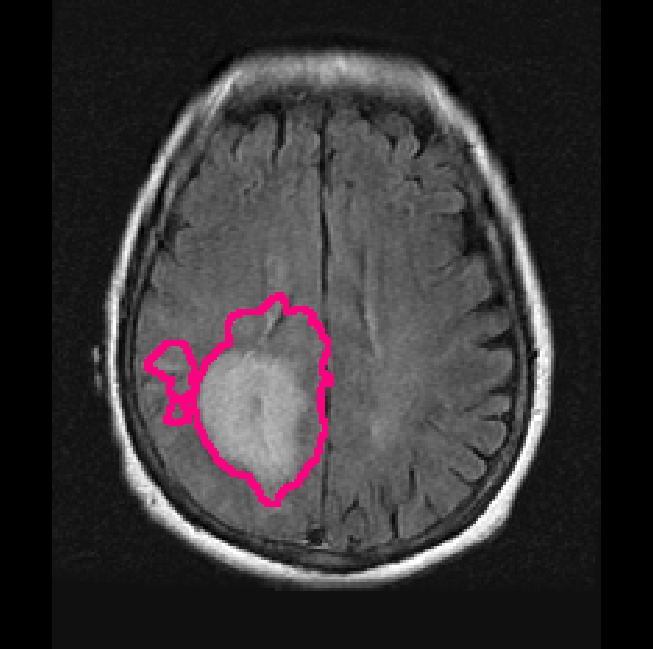}
        \end{subfigure}
        \caption{Patient TCGA-02-0037 from the TCGA-GBM data collection.}
    \end{subfigure}
    \begin{subfigure}[b]\textwidth
        \centering
        \hfill
        \begin{subfigure}[b]{0.215\textwidth}
        \includegraphics[width=\textwidth, clip, trim=2.5cm 0.5cm 2.5cm 0.5cm]{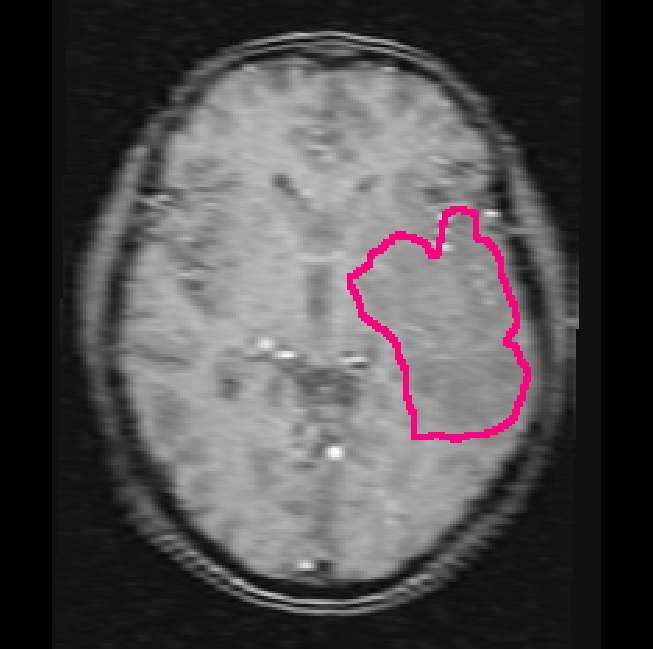}
        \end{subfigure}
        \hfill
        \begin{subfigure}[b]{0.215\textwidth}
        \includegraphics[width=\textwidth, clip, trim=2.5cm 0.5cm 2.5cm 0.5cm]{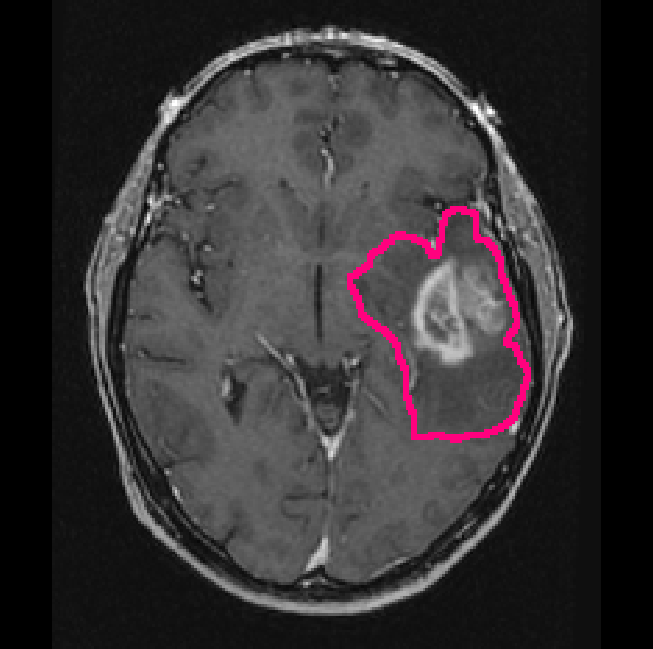}
        \end{subfigure}
        \hfill
        \begin{subfigure}[b]{0.215\textwidth}
        \includegraphics[width=\textwidth, clip, trim=2.5cm 0.5cm 2.5cm 0.5cm]{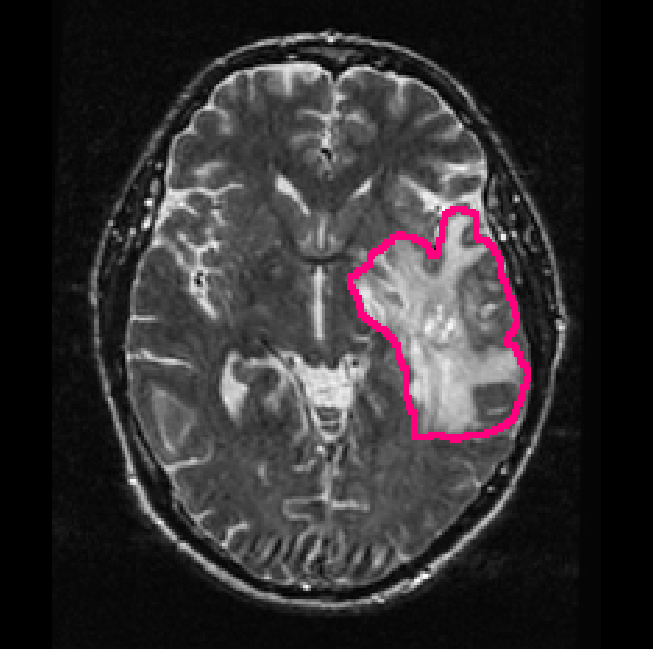}
        \end{subfigure}
        \hfill
        \begin{subfigure}[b]{0.215\textwidth}
        \includegraphics[width=\textwidth, clip, trim=2.5cm 0.5cm 2.5cm 0.5cm]{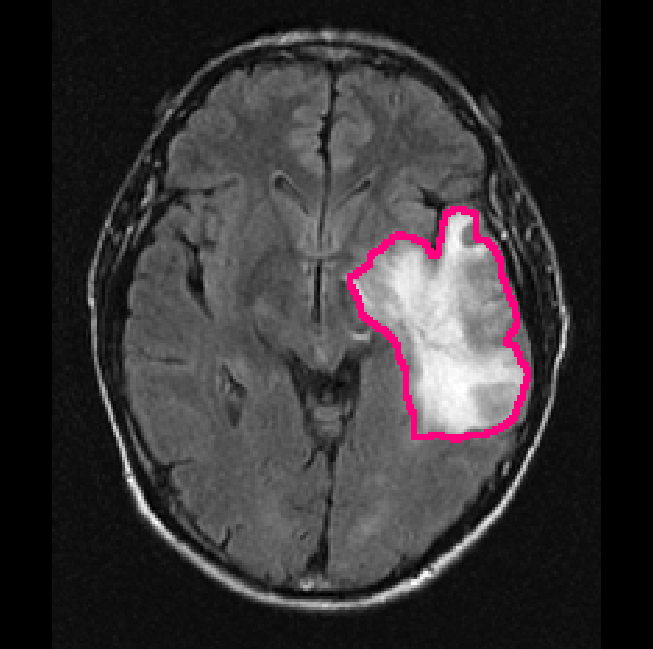}
        \end{subfigure}
        \caption{Patient TCGA-08-0353 from the TCGA-GBM data collection.}
    \end{subfigure}
    \begin{subfigure}[b]\textwidth
        \centering
        \hfill
        \begin{subfigure}[b]{0.215\textwidth}
        \includegraphics[width=\textwidth, clip, trim=2.5cm 0.5cm 2.5cm 0.5cm]{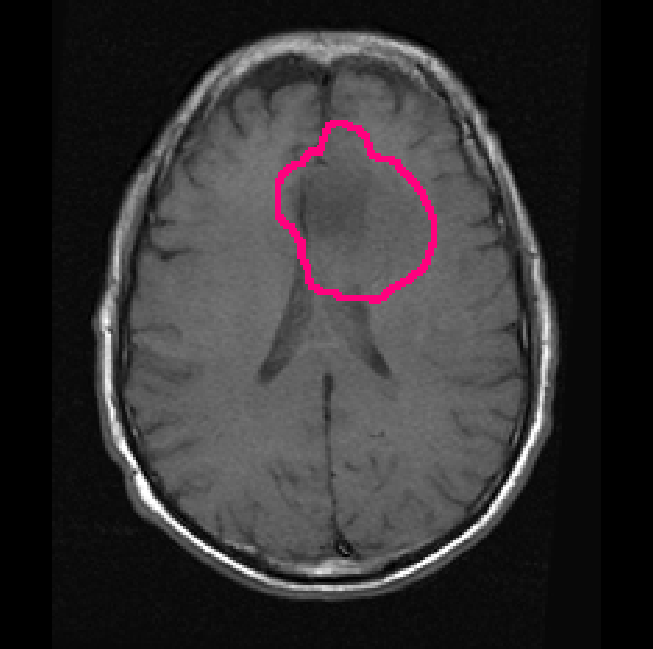}
        \end{subfigure}
        \hfill
        \begin{subfigure}[b]{0.215\textwidth}
        \includegraphics[width=\textwidth, clip, trim=2.5cm 0.5cm 2.5cm 0.5cm]{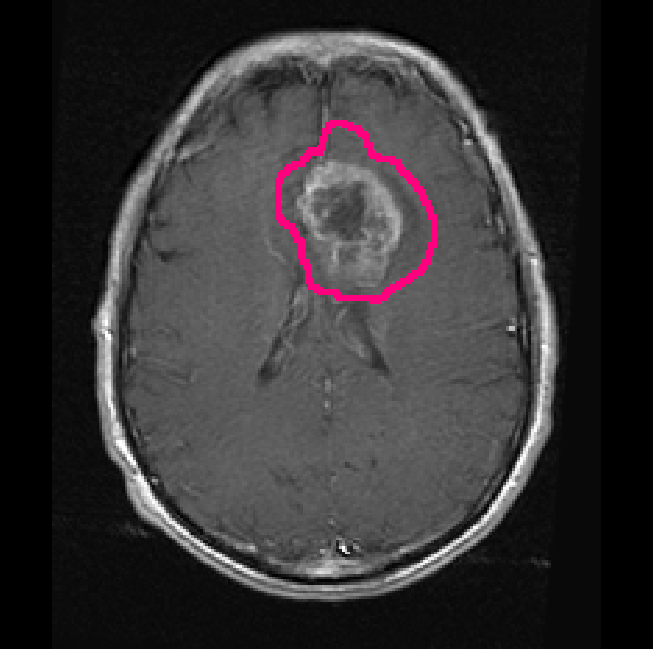}
        \end{subfigure}
        \hfill
        \begin{subfigure}[b]{0.215\textwidth}
        \includegraphics[width=\textwidth, clip, trim=2.5cm 0.5cm 2.5cm 0.5cm]{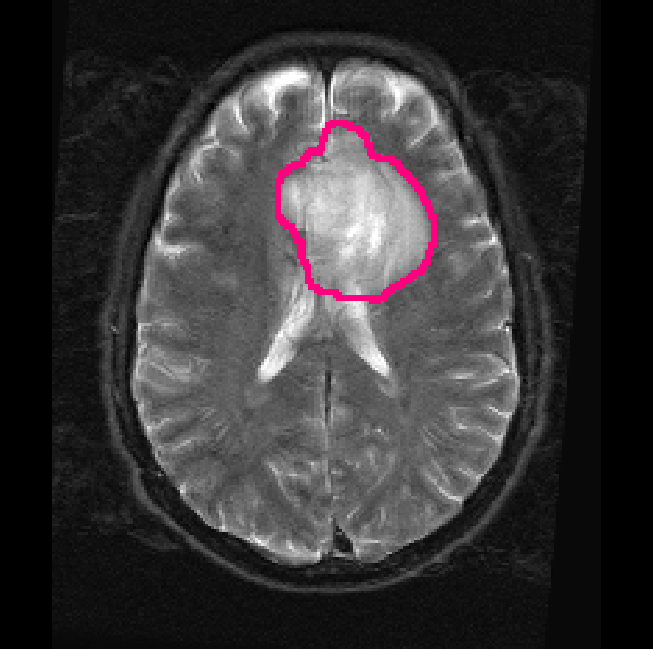}
        \end{subfigure}
        \hfill
        \begin{subfigure}[b]{0.215\textwidth}
        \includegraphics[width=\textwidth, clip, trim=2.5cm 0.5cm 2.5cm 0.5cm]{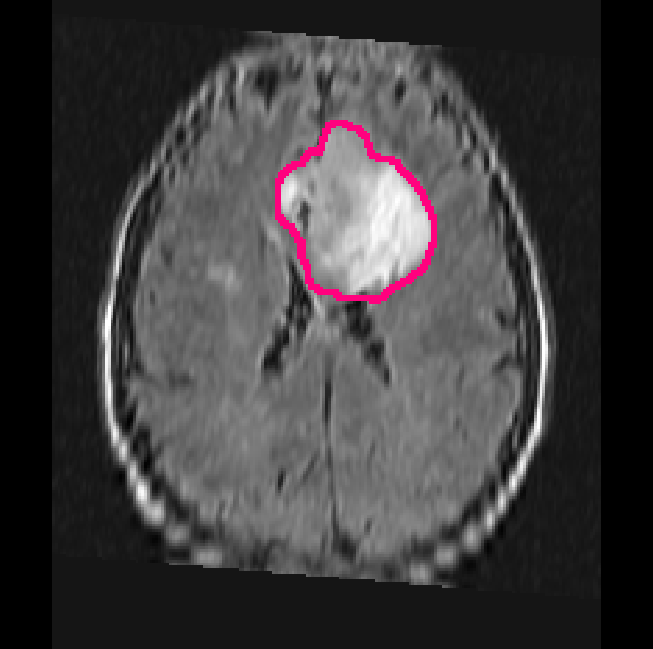}
        \end{subfigure}
        \caption{Patient TCGA-12-1094 from the TCGA-GBM data collection.}
    \end{subfigure}
    \begin{subfigure}[b]\textwidth
        \centering
        \hfill
        \begin{subfigure}[b]{0.215\textwidth}
        \includegraphics[width=\textwidth, clip, trim=2.5cm 0.5cm 2.5cm 0.5cm]{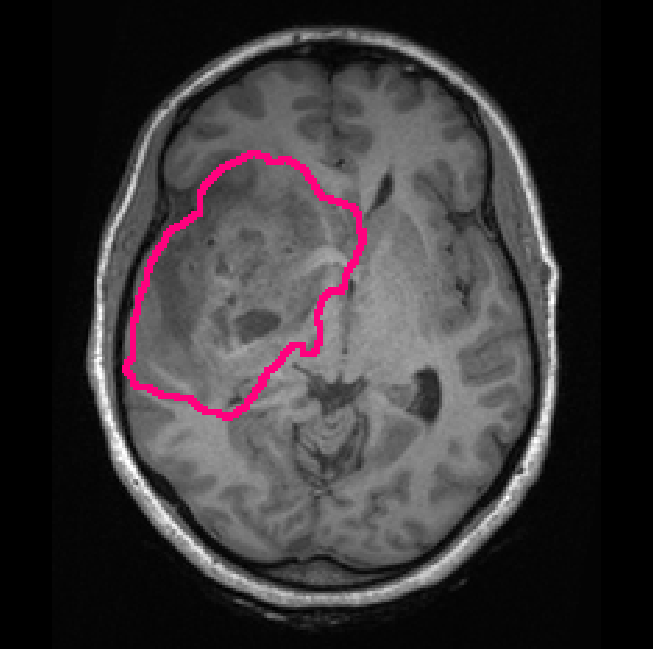}
        \end{subfigure}
        \hfill
        \begin{subfigure}[b]{0.215\textwidth}
        \includegraphics[width=\textwidth, clip, trim=2.5cm 0.5cm 2.5cm 0.5cm]{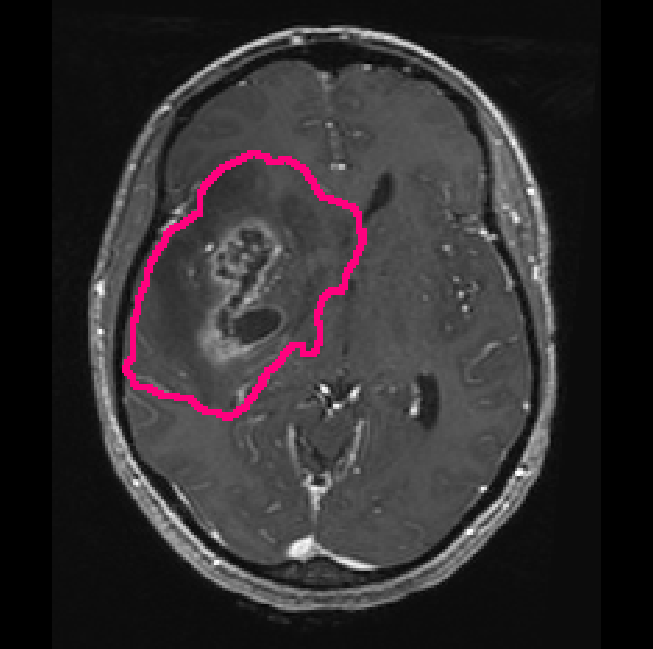}
        \end{subfigure}
        \hfill
        \begin{subfigure}[b]{0.215\textwidth}
        \includegraphics[width=\textwidth, clip, trim=2.5cm 0.5cm 2.5cm 0.5cm]{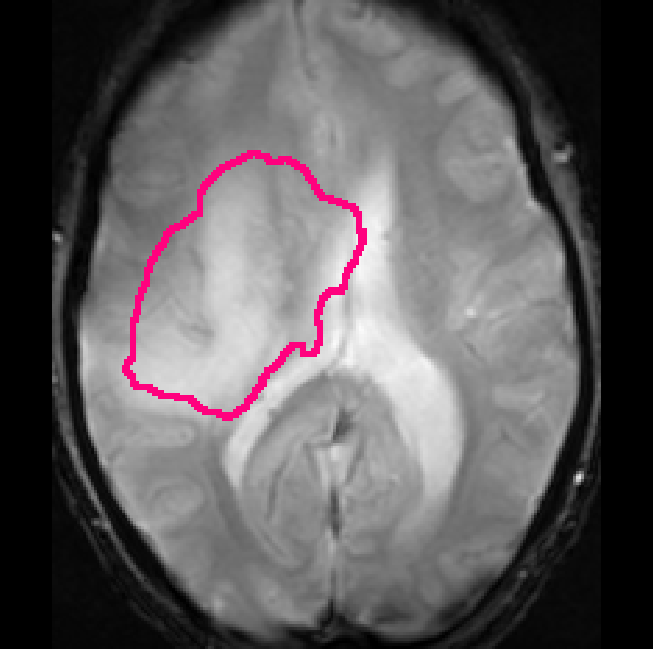}
        \end{subfigure}
        \hfill
        \begin{subfigure}[b]{0.215\textwidth}
        \includegraphics[width=\textwidth, clip, trim=2.5cm 0.5cm 2.5cm 0.5cm]{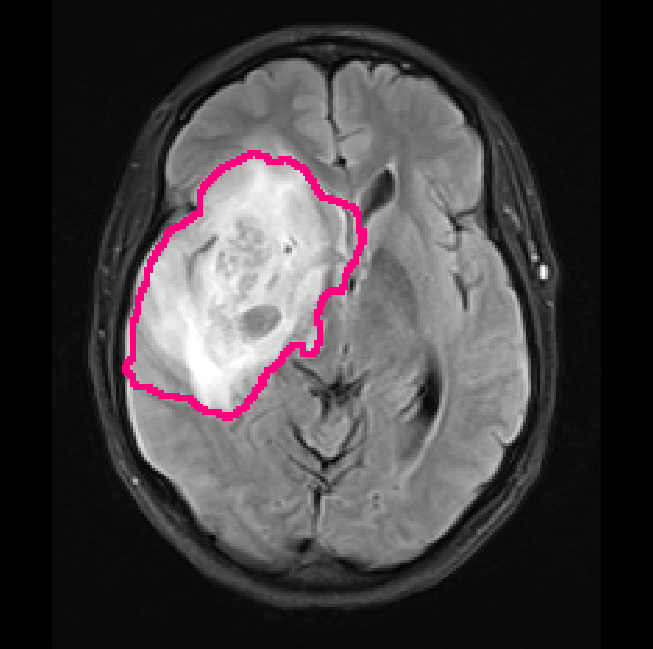}
        \end{subfigure}
        \caption{Patient TCGA-14-3477 from the TCGA-GBM data collection.}\label{fig:hgg_example_tcga143477}
    \end{subfigure}
    \begin{subfigure}[b]\textwidth
        \centering
        \hfill
        \begin{subfigure}[b]{0.215\textwidth}
        \includegraphics[width=\textwidth, clip, trim=2.5cm 0.5cm 2.5cm 0.5cm]{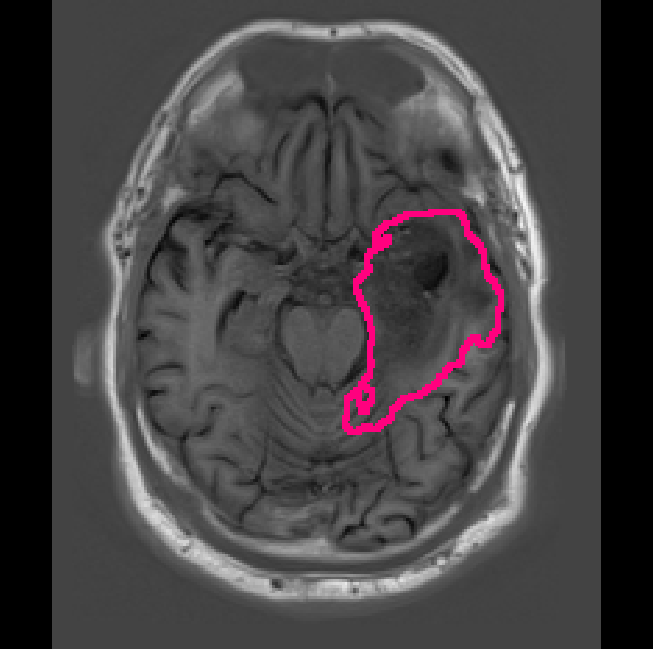}
        \end{subfigure}
        \hfill
        \begin{subfigure}[b]{0.215\textwidth}
        \includegraphics[width=\textwidth, clip, trim=2.5cm 0.5cm 2.5cm 0.5cm]{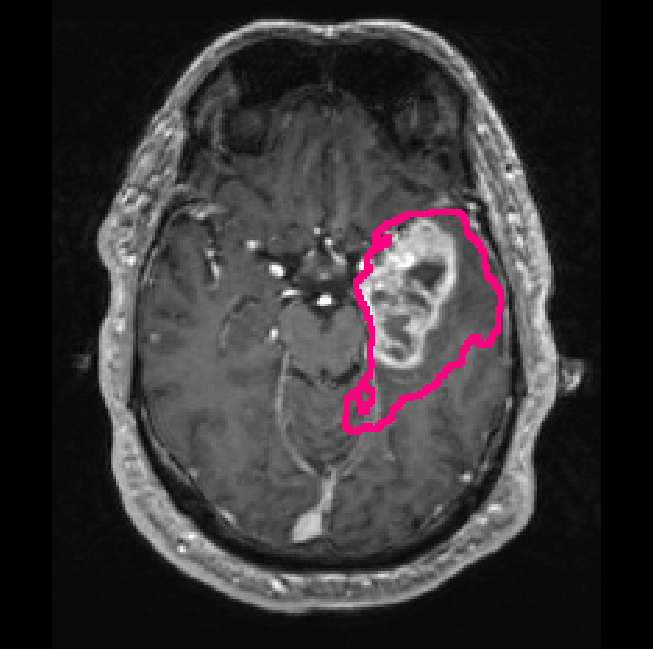}
        \end{subfigure}
        \hfill
        \begin{subfigure}[b]{0.215\textwidth}
        \includegraphics[width=\textwidth, clip, trim=2.5cm 0.5cm 2.5cm 0.5cm]{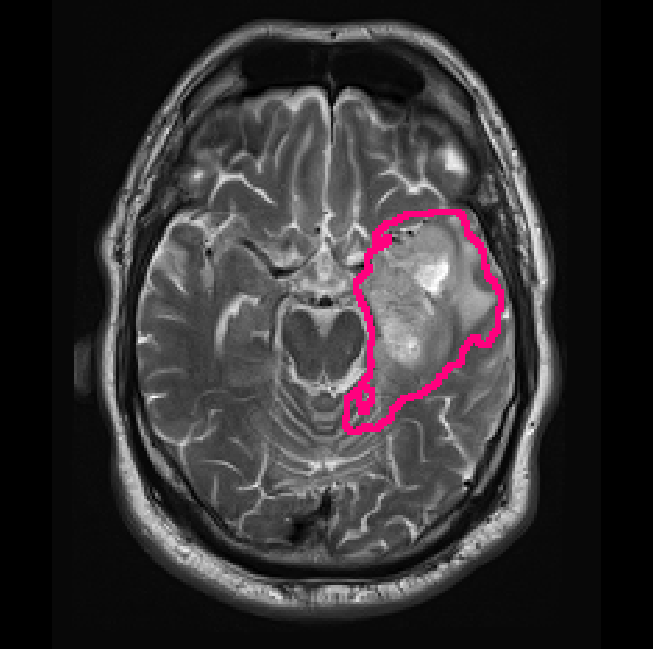}
        \end{subfigure}
        \hfill
        \begin{subfigure}[b]{0.215\textwidth}
        \includegraphics[width=\textwidth, clip, trim=2.5cm 0.5cm 2.5cm 0.5cm]{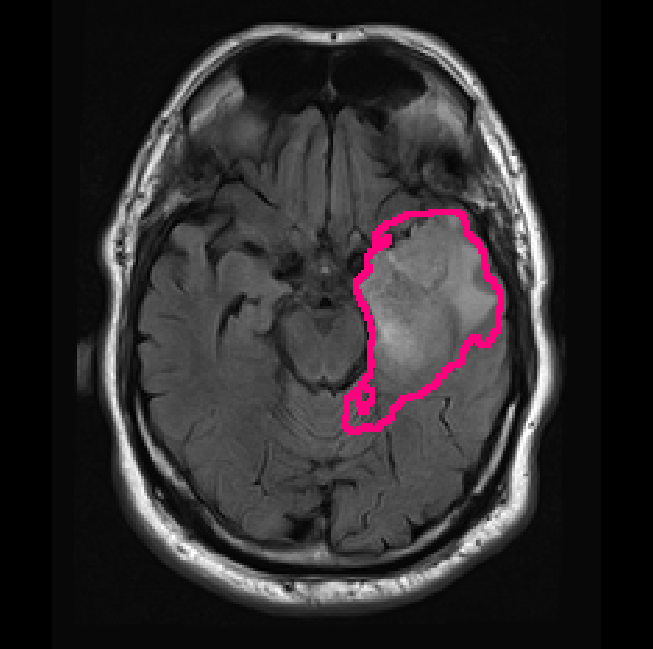}
        \end{subfigure}
        \caption{Patient TCGA-19-5951 from the TCGA-GBM data collection.}
    \end{subfigure}
    \caption{Examples of scans and automatic segmentations of five patients that were randomly selected from the TCGA-GBM data collection.}\label{fig:seg_examples_HGG}
\end{figure}

To demonstrate the automatic segmentations made by our method, we randomly selected five patients from both the \gls{TCGALGG} and the \gls{TCGAGBM} dataset.
The scans and segmentations of the five patients from the \gls{TCGALGG} dataset and the \gls{TCGAGBM} dataset are shown in \cref{fig:seg_examples_LGG,fig:seg_examples_HGG}, respectively.
The DICE score, Hausdorff distance, and volumetric similarity coefficient for these patients are given in \cref{tab:seg_metrics_example_patients}.
The method seems to mostly focus on the hyperintensities of the \gls{FLAIR} scan.
Despite the registrations issues that can be seen for the \gls{T2} scan in \cref{fig:hgg_example_tcga143477} the tumor was still properly segmented, demonstrating the robustness of our method.

\begin{table}[htbp]
\begin{tabular}{lC{2cm}C{2cm}C{2cm}}
    \toprule
    Patient & DICE & HD (mm)& VSC\\
    \midrule
    TCGA-LGG\\
    \cmidrule(lr){1-1}
    TCGA-DU-7301 & 0.89 & 10.3 & 0.95\\
    TCGA-FG-5964 & 0.80 & \hphantom{0}5.8 & 0.82\\
    TCGA-FG-A713 & 0.73 & \hphantom{0}7.8 & 0.88\\
    TCGA-HT-7475 & 0.87 & 14.9 & 0.90\\
    TCGA-HT-8106 & 0.88 & 11.2 & 0.99\\
    \\
    TCGA-GBM\\
    \cmidrule(lr){1-1}
    TCGA-02-0037 & 0.82 & 22.6 & 0.99\\
    TCGA-08-0353 & 0.91 & 13.0 & 0.98\\
    TCGA-12-1094 & 0.90 & \hphantom{0}7.3 & 0.93\\
    TCGA-14-3477 & 0.90 & 16.5 & 0.99\\
    TCGA-19-5951 & 0.73 & 19.7 & 0.73\\
    \bottomrule
\end{tabular}
\caption{The DICE score, Hausdorff distance (HD), and volumetric similarity coefficient (VSC) for the randomly selected patients from the TCGA-LGG and TCGA-GBM data collections.}\label{tab:seg_metrics_example_patients}
\end{table}

\clearpage
\newpage

\section{Prediction results in the test set}\label{app:individual_pred}
\underline{\textattachfile[color=0 0 0.54,mimetype=text/csv]{Appendices/test_predictions.csv}{Supplementary file}}

\newpage

\section{Filter output visualizations}\label{app:filter_vis}

\begin{figure}
    \begin{subfigure}[b]{0.9\textwidth}
        \centering
        \hfill
        \begin{subfigure}[b]{0.24\textwidth}
        \includegraphics[width=\textwidth]{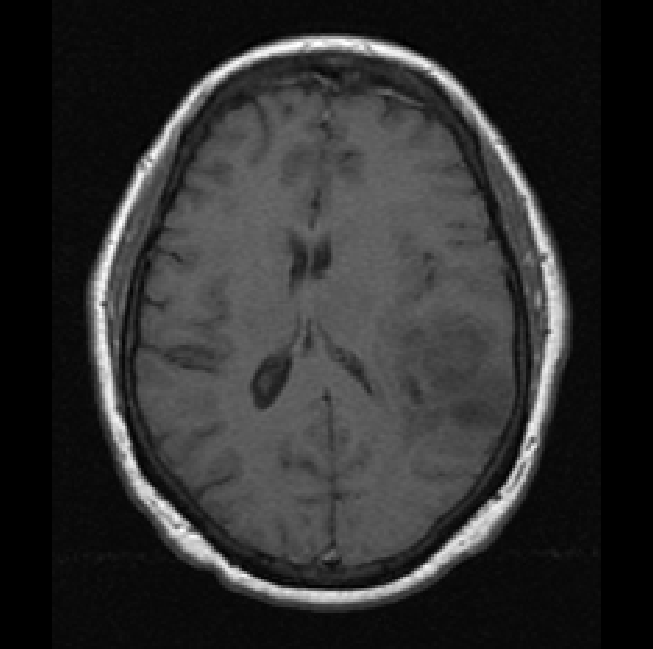}
        \caption*{Pre-contrast \acrshort{T1}}
        \end{subfigure}
        \hfill
        \begin{subfigure}[b]{0.24\textwidth}
        \includegraphics[width=\textwidth]{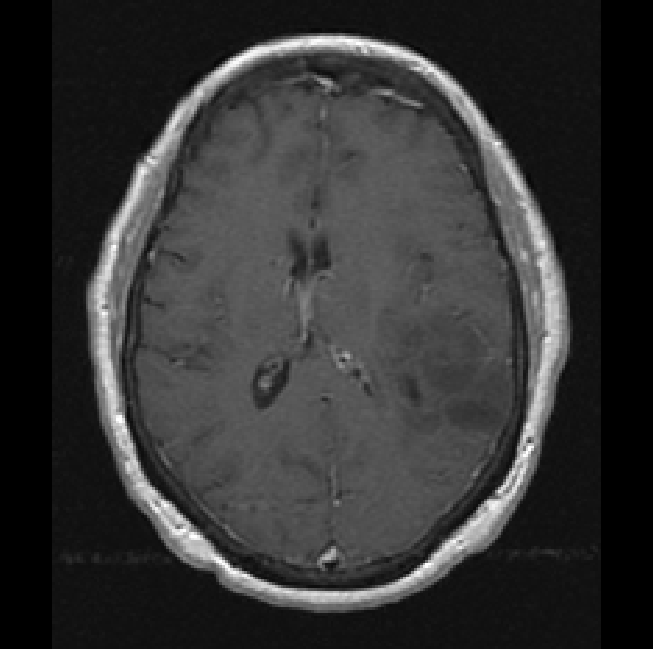}
        \caption*{Post-contrast \acrshort{T1}}
        \end{subfigure}
        \hfill
        \begin{subfigure}[b]{0.24\textwidth}
        \includegraphics[width=\textwidth]{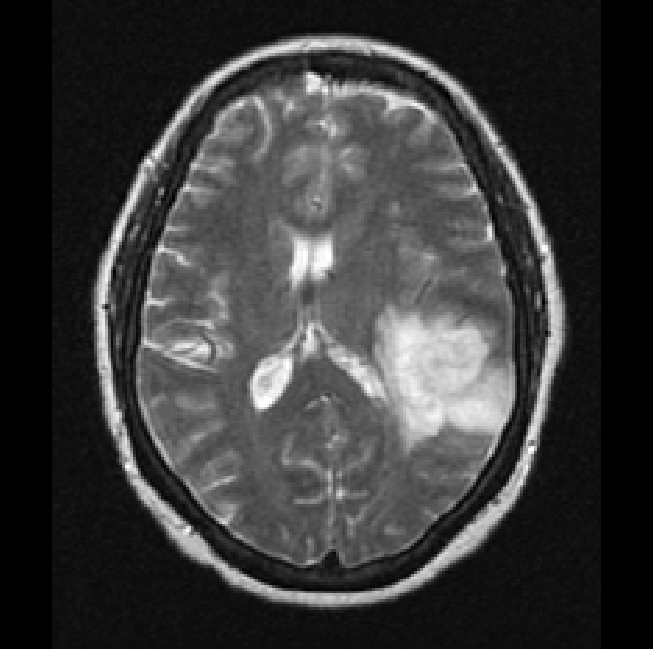}
        \caption*{\acrshort{T2}}
        \end{subfigure}
        \hfill
        \begin{subfigure}[b]{0.24\textwidth}
        \includegraphics[width=\textwidth]{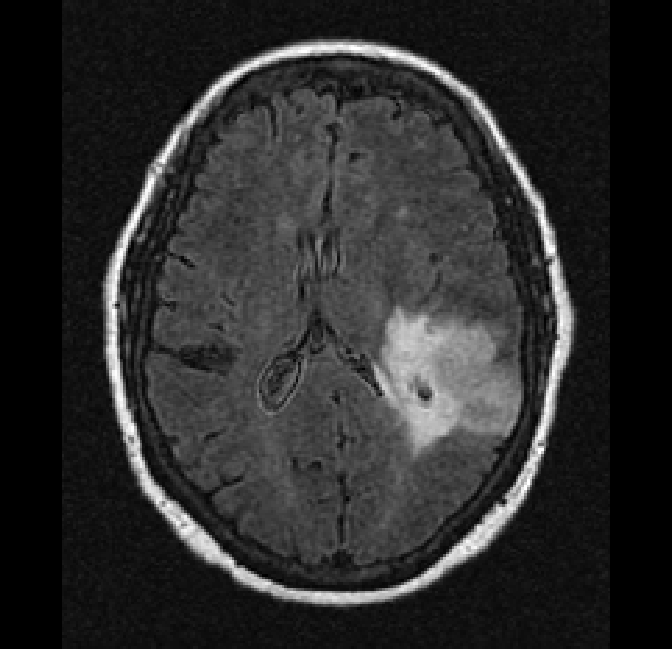}
        \caption*{\acrshort{FLAIR}}
        \end{subfigure}
        \caption{Scans used to derive the convolutional layer filter output visualizations.}
    \end{subfigure}
    \vskip\baselineskip
    \begin{subfigure}[b]{0.9\textwidth}
        \centering
        \includegraphics[width=\textwidth]{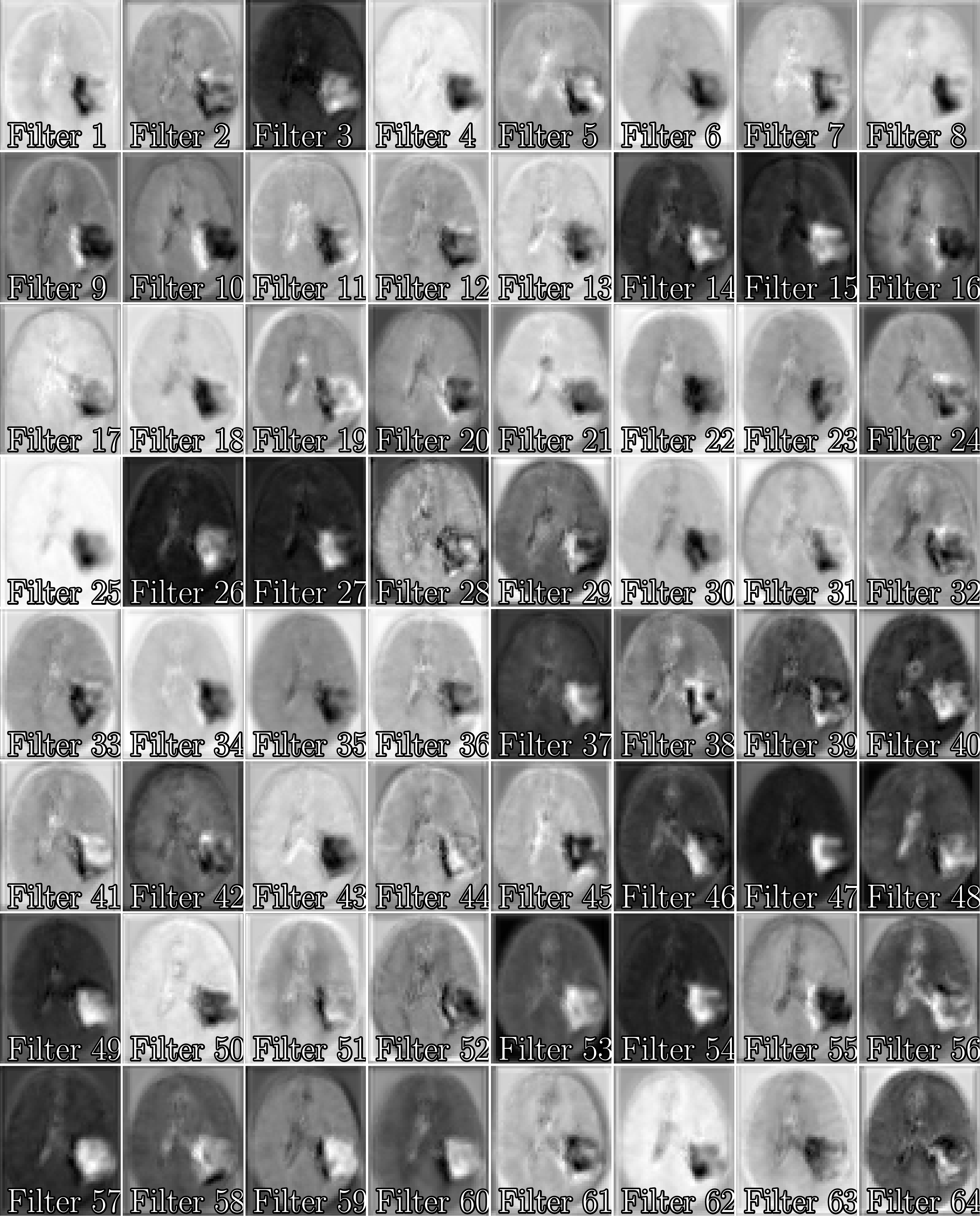}
        \caption{Filter output visualizations.}
    \end{subfigure}
    \caption{Filter output visualizations of the last convolutional layer in the downsample path of the network at feature map size 49x61x51 for patient TCGA-DU-6400.
    This is an IDH mutated,  1p/19q co-deleted, grade II glioma.}\label{fig:filter_lgg_shallow}
\end{figure}

\begin{figure}
    \begin{subfigure}[b]{0.9\textwidth}
        \centering
        \hfill
        \begin{subfigure}[b]{0.24\textwidth}
        \includegraphics[width=\textwidth]{Figures/TCGA-DU-6400_T1.png}
        \caption*{Pre-contrast \acrshort{T1}}
        \end{subfigure}
        \hfill
        \begin{subfigure}[b]{0.24\textwidth}
        \includegraphics[width=\textwidth]{Figures/TCGA-DU-6400_T1GD.png}
        \caption*{Post-contrast \acrshort{T1}}
        \end{subfigure}
        \hfill
        \begin{subfigure}[b]{0.24\textwidth}
        \includegraphics[width=\textwidth]{Figures/TCGA-DU-6400_T2.png}
        \caption*{\acrshort{T2}}
        \end{subfigure}
        \hfill
        \begin{subfigure}[b]{0.24\textwidth}
        \includegraphics[width=\textwidth]{Figures/TCGA-DU-6400_FLAIR.png}
        \caption*{\acrshort{FLAIR}}
        \end{subfigure}
        \caption{Scans used to derive the convolutional layer filter output visualizations.}
    \end{subfigure}
    \vskip\baselineskip
    \begin{subfigure}[b]{0.9\textwidth}
        \centering
        \includegraphics[width=\textwidth]{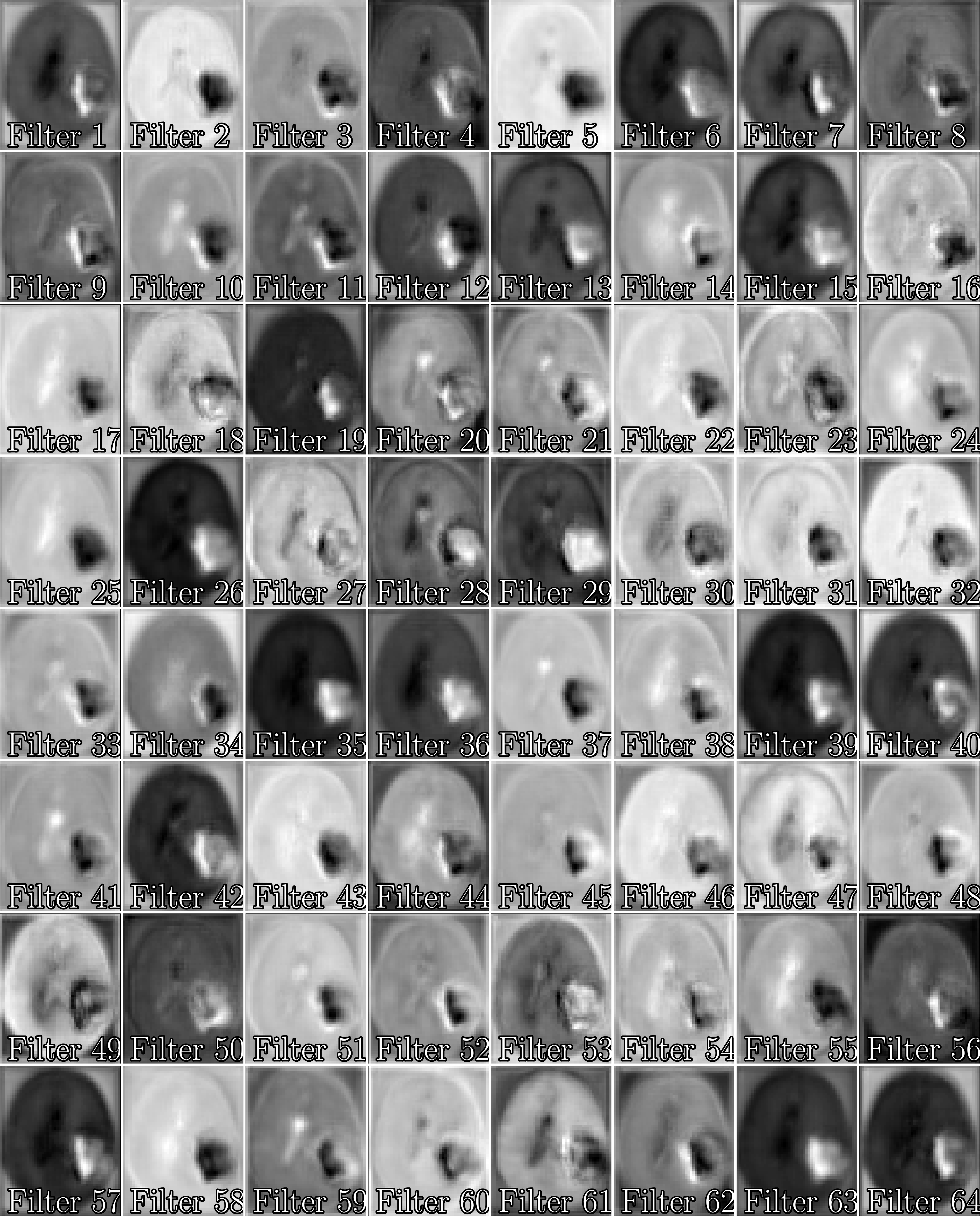}
        \caption{Filter output visualizations.}
    \end{subfigure}
    \caption{Filter output visualizations of the last convolutional layer in the upsample path of the network at feature map size 49x61x51 for patient TCGA-DU-6400.
    This is an IDH mutated, 1p/19q co-deleted, grade II glioma.}\label{fig:filter_lgg_deep}
\end{figure}

\begin{figure}
    \begin{subfigure}[b]{0.9\textwidth}
        \centering
        \hfill
        \begin{subfigure}[b]{0.24\textwidth}
        \includegraphics[width=\textwidth]{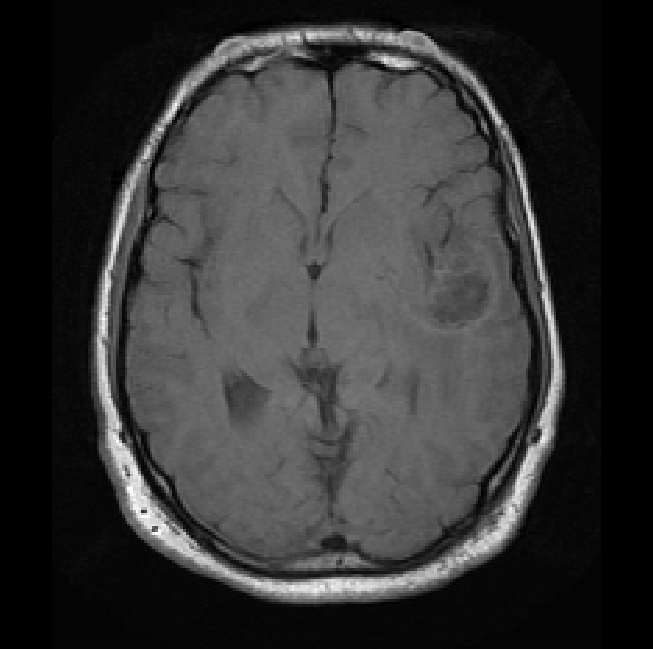}
        \caption*{Pre-contrast \acrshort{T1}}
        \end{subfigure}
        \hfill
        \begin{subfigure}[b]{0.24\textwidth}
        \includegraphics[width=\textwidth]{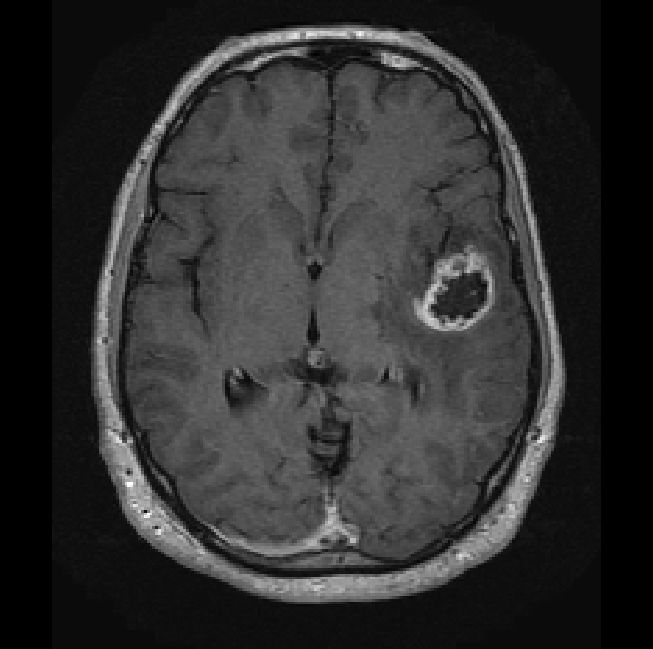}
        \caption*{Post-contrast \acrshort{T1}}
        \end{subfigure}
        \hfill
        \begin{subfigure}[b]{0.24\textwidth}
        \includegraphics[width=\textwidth]{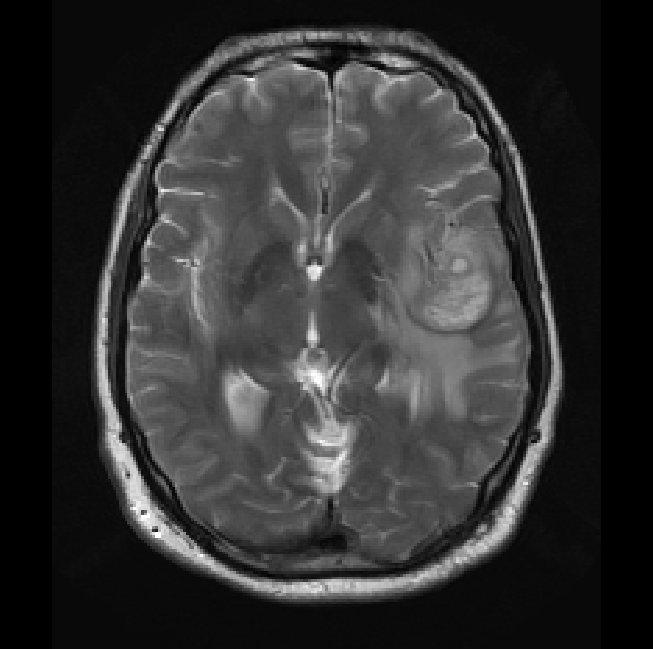}
        \caption*{\acrshort{T2}}
        \end{subfigure}
        \hfill
        \begin{subfigure}[b]{0.24\textwidth}
        \includegraphics[width=\textwidth]{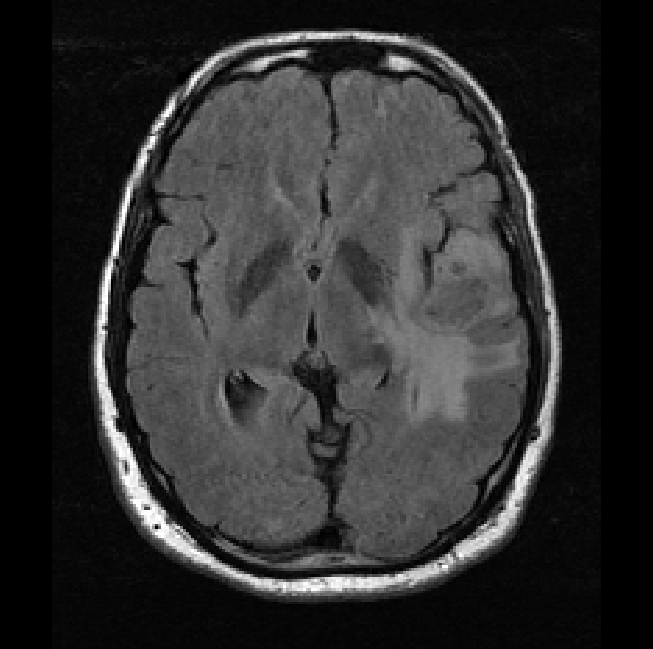}
        \caption*{\acrshort{FLAIR}}
        \end{subfigure}
        \caption{Scans used to derive the convolutional layer filter output visualizations.}
    \end{subfigure}
    \begin{subfigure}[b]{0.9\textwidth}
        \centering
        \includegraphics[width=\textwidth]{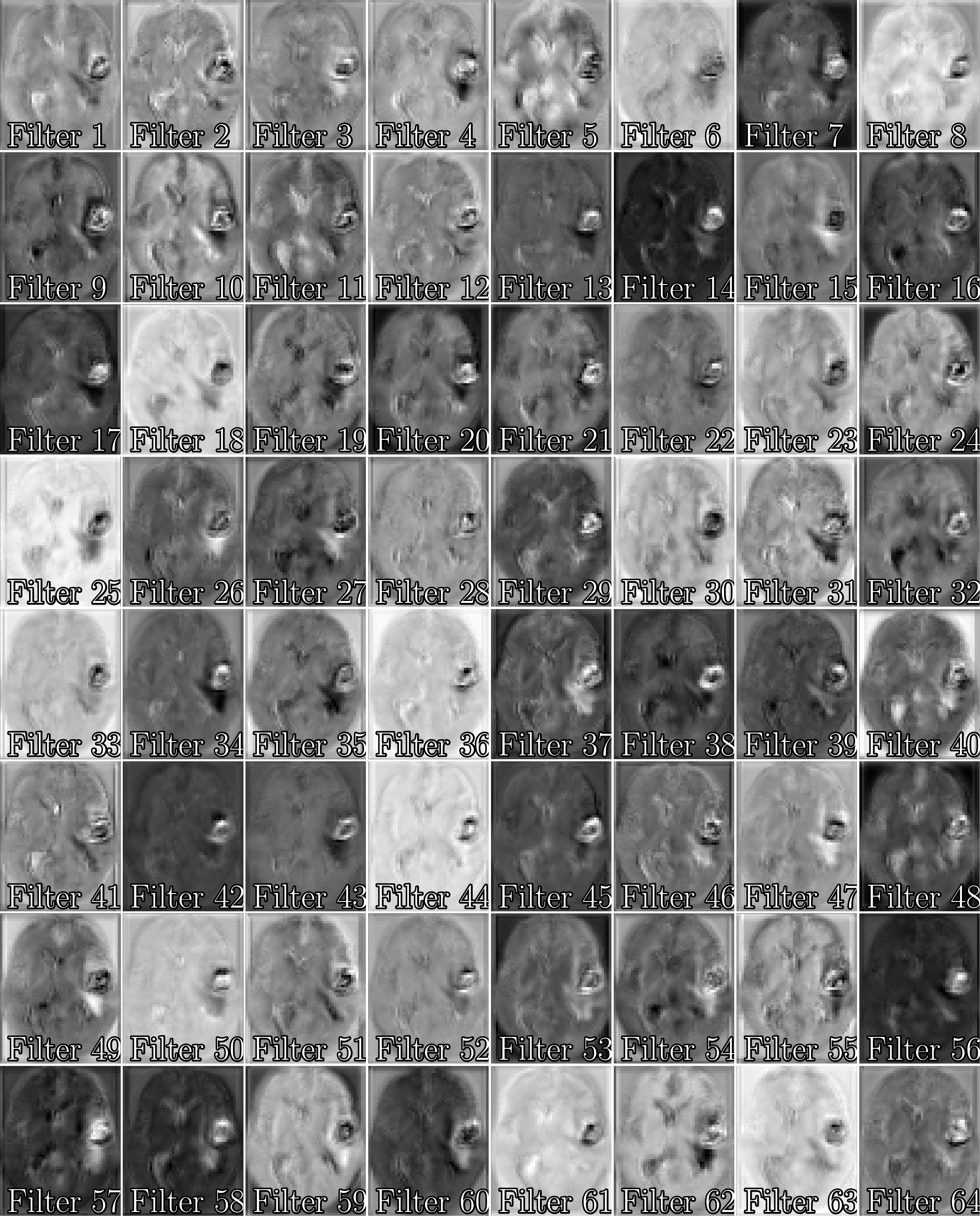}
        \caption{Filter output visualizations.}
    \end{subfigure}
    \caption{Filter output visualizations of the last convolutional layer in the downsample path of the network at feature map size 49x61x51 for patient TCGA-06-0238.
    This is an IDH wildtype, grade IV glioma.}\label{fig:filter_hgg_shallow}
\end{figure}

\begin{figure}
    \begin{subfigure}[b]{0.9\textwidth}
        \centering
        \hfill
        \begin{subfigure}[b]{0.24\textwidth}
        \includegraphics[width=\textwidth]{Figures/TCGA-06-0238_T1.png}
        \caption*{Pre-contrast \acrshort{T1}}
        \end{subfigure}
        \hfill
        \begin{subfigure}[b]{0.24\textwidth}
        \includegraphics[width=\textwidth]{Figures/TCGA-06-0238_T1GD.png}
        \caption*{Post-contrast \acrshort{T1}}
        \end{subfigure}
        \hfill
        \begin{subfigure}[b]{0.24\textwidth}
        \includegraphics[width=\textwidth]{Figures/TCGA-06-0238_T2.png}
        \caption*{\acrshort{T2}}
        \end{subfigure}
        \hfill
        \begin{subfigure}[b]{0.24\textwidth}
        \includegraphics[width=\textwidth]{Figures/TCGA-06-0238_FLAIR.png}
        \caption*{\acrshort{FLAIR}}
        \end{subfigure}
        \caption{Scans used to derive the convolutional layer filter output visualizations.}
    \end{subfigure}
    \begin{subfigure}[b]{0.9\textwidth}
        \centering
        \includegraphics[width=\textwidth]{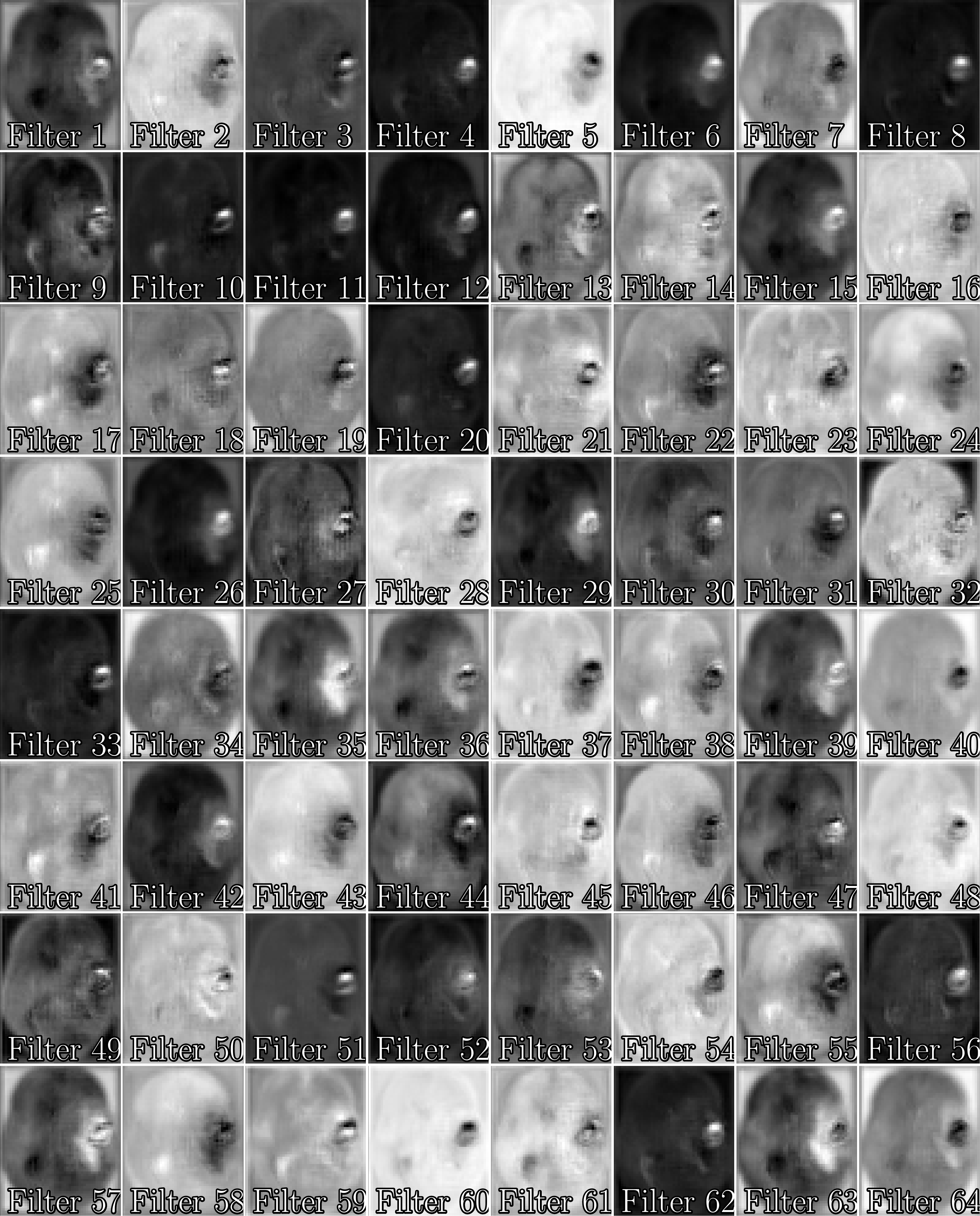}
        \caption{Filter output visualizations.}
    \end{subfigure}
    \caption{Filter output visualizations of the last convolutional layer in the upsample path of the network at feature map size 49x61x51 for patient TCGA-06-0238.
     This is an IDH wildtype, grade IV glioma.}\label{fig:filter_hgg_deep}
\end{figure}

\cref{fig:filter_lgg_shallow,fig:filter_lgg_deep} show the output of the convolution filters for the same \gls{LGG} patient as shown in \cref{fig:saliency_LGG}, and \cref{fig:filter_hgg_shallow,fig:filter_hgg_deep}  show the output of the convolution filters for the same \gls{HGG} patient as shown in \cref{fig:saliency_HGG}.
\cref{fig:filter_lgg_shallow,fig:filter_hgg_shallow} show the outputs of the last convolution layer in the downsample path at the feature size of 49x61x51 (the fourth convolutional layer in the network).
\cref{fig:filter_lgg_deep,fig:filter_hgg_deep} show the outputs of the last convolution layer in the upsample path at the feature size of 49x61x51 (the nineteenth convolutional layer in the network).

Comparing \cref{fig:filter_lgg_shallow} to \cref{fig:filter_lgg_deep} and \cref{fig:filter_hgg_shallow} to \cref{fig:filter_hgg_deep} we can see that the convolutional layers in the upsample path do not keep a lot of detail for the healthy part of the brain, as this region seems blurred.
However, within the tumor different regions can still be distinguished.
The different parts of the tumor from the scans can also be seen, such as the contrast-enhancing part and the high signal intensity on the \gls{FLAIR}.
For the grade IV glioma in \cref{fig:filter_hgg_deep}, some filters, such as filter 26, also seem to focus on the necrotic part of the tumor.

\newpage

\section{Training losses}\label{app:losses}

During the training of the network we used masked categorical cross-entropy loss for the IDH, 1p/19q, and grade outputs.
The normal categorical cross-entropy loss is defined as:

\begin{equation}
    \mathcal{L}^{CE}_{batch} = -\frac{1}{N_{batch}} \sum_{j} \sum_{i \in C} y_{i, j} \log \left(\hat{y}_{i, j} \right),
\label{eq:ce_loss}
\end{equation}
where $\mathcal{L}^{CE}_{batch}$ is the total cross-entropy loss over a batch, $y_{i,j}$ is the ground truth label of sample $j$ for class $i$, $\hat{y}_{i,j}$ is the prediction score for sample $j$ for class $i$, $C$ is the set of classes, and $N_{batch}$ is the number of samples in the batch.
Here it is assumed that the ground truth labels are one-hot-encoded, thus $y_{i,j}$ is either 0 or 1 for each class.
In our case, the ground truth is not known for all samples, which can be incorporated in  \cref{eq:ce_loss} by setting $y_{i,j}$ to 0 for all classes for a sample for which the ground truth is not known.
That sample would then not contribute to the overall loss, and would not contribute to the gradient update.
However, this can skew the total loss over a batch, since the loss is still averaged over the total number of samples in a batch, regardless of whether the ground truth is known, resulting in a lower loss for batches that contained more samples with unknown ground truth.
Therefore, we used a masked categorical cross-entropy loss:

\begin{equation}
    \mathcal{L}^{CE}_{batch} = -\frac{1}{N_{batch}} \sum_{j} \mu_j^{batch} \sum_{i \in C} y_{i, j} \log \left(\hat{y}_{i, j} \right),
\end{equation}
where
\begin{equation}
\mu_j^{batch} = \frac{N_{batch}}{\sum_{i,j} y_{i,j}} \sum_i y_{i,j}
\end{equation}
is the batch weight for sample $j$.
In this way, the total batch loss is only averaged over the samples that actually have a ground truth.

Since there was an imbalance between the number of ground truth samples for each class, we used class weights to compensate for this imbalance.
Thus, the loss becomes:

\begin{equation}
    \mathcal{L}^{CE}_{batch} = -\frac{1}{N_{batch}} \sum_{j} \mu_j^{batch} \sum_{i \in C} \mu_i^{class} y_{i, j} \log \left(\hat{y}_{i, j} \right),
\end{equation}
where
\begin{equation}
    \mu_{i}^{class} =  \frac{N}{N_{i} \left|C\right|}
\end{equation}
is the class weight for class $i$, $N$ is the total number of samples with known ground truth, $N_i$ is the number of samples of class $i$, and $\left|C\right|$ is the number of classes.
By determining the class weight in this way, we ensured that:

\begin{equation}
    \mu_i^{class} N_{i} = \frac{N}{\left|C\right|} = \text{constant}.
\end{equation}
Thus, each class would have the same contribution to the overall loss.
These class weights were (individually) determined for the IDH output, the 1p/19q output, and the grade output.

For the segmentation output we used the DICE loss:

\begin{equation}
    \mathcal{L}^{DICE}_{batch} = \sum_j 1 - 2 \cdot \frac{\sum_k^{voxels} y_{j, k} \cdot \hat{y}_{j, k}}{\sum_k^{voxels} y_{j, k} + \hat{y}_{j, k}},
\end{equation}
where $y_{j, k}$ is the ground truth label in voxel $k$ of sample $j$, and $\hat{y}_{j,k}$ is the prediction score outputted for voxel $k$ of sample $j$.

The total loss that was optimized for the model was a weighted sum of the four individual losses:

\begin{equation}
    \mathcal{L}^{total} = \sum_m \mu_m \mathcal{L}_m,
\end{equation}
with
\begin{equation}
    \mu_m = \frac{1}{X_m},
\end{equation}
where $\mathcal{L}_m$ is the loss for output $m$, $\mu_m$ is the loss weight for loss $m$ (either the IDH, 1p/19q, grade or segmentation loss), and $X_m$ is the number of samples with known ground truth for output $m$.
In this way, we could counteract the effect of certain outputs having more known labels than other outputs.

\newpage

\section{Parameter tuning}\label{app:hyperparameter_tuning}

\begin{table}[htbp]
\centering
\caption{Hyperparameters that were tuned, and the values that were tested. Values in bold show the selected values used in the final model.}
\begin{tabular}{ll}
\toprule
Tuning parameter & Tested values\\
\midrule
Dropout rate & 0.15, 0.2, \textbf{0.25}, 0.30, 0.35, 0.40\\
$l2$-norm & 0.0001, \textbf{0.00001}, 0.000001\\
Learning rate & 0.01, 0.001, 0.0001, \textbf{0.00001}, 0.0000001\\
Weight decay & 0.001, 0.0001, \textbf{0.00001}\\
Augmentation factor & 1, \textbf{2}, 3\\
Augmentation probability & 0.25, 0.30, \textbf{0.35}, 0.40, 0.45\\
\bottomrule
\end{tabular}
\end{table}

\newpage
\section{Evaluation metrics}\label{app:metric_defs}

We calculated the AUC, accuracy, sensitivity, and specificity metrics for the genetic and histological features; for the definitions of these metrics see \cite{tharwat2018metrics}.

For the \gls{IDH} and 1p/19q co-deletion outputs, the \gls{IDH} mutated and the 1p/19q co-deleted samples were regarded as the positive class respectively.
Since the grade was a multi-class problem, no single positive class could be determined.
For the prediction of the individual grades, that grade was seen as the positive class and all other grades as the negative class (e.g., in the case of the grade III prediction, grade III was regarded as the positive class, and grade II and IV were regarded as the negative class).
For the \gls{LGG} vs. \gls{HGG} prediction, \gls{LGG} was considered as the positive class and \gls{HGG} as the negative class.
For the evaluation of these metrics for the genetic and histological features, only the subjects with known ground truth were taken into account.

The overall AUC for the grade was a multi-class AUC determined in a one-vs-one approach, comparing each class against the others; in this way, this metric was insensitive to class imbalance \cite{hand2001AUC}.
A multi-class accuracy was used to determine the overall accuracy for the grade predictions \cite{tharwat2018metrics}.

To evaluate the performance of the automated segmentation, we  evaluated the DICE score, the Hausdorff distance and the volumetric similarity coefficient.
The DICE score is a measure of overlap between two segmentations, where a value of 1 indicates perfect overlap, and the Hausdorff distance is a measure of the closeness of the borders of the segmentations.
The volumetric similarity coefficient is a measure of the agreement between the volumes of two segmentations, without taking account the actual location of the tumor, where a value of 1 indicates perfect agreement.
See \cite{taha2015metrics} for the definitions of these metrics.

\newpage

\section{Ground truth labels of patients included from public datasets}\label{app:open_gt}
\underline{\textattachfile[color=0 0 0.54,mimetype=text/csv]{Appendices/included_patient_labels.csv}{Supplementary file}}

\section*{Acknowledgments}

Sebastian van der Voort and Fatih Incekara acknowledge funding by the Dutch Cancer Society (KWF project number EMCR 2015-7859).

Data used in this publication were generated by the National Cancer Institute Clinical Proteomic Tumor Analysis Consortium (CPTAC).

The results published here are in whole or part based upon data generated by the TCGA Research Network: \url{http://cancergenome.nih.gov/}.

\section*{Author contributions}

S.R.v.d.V., F.I., W.J.N., M.S., and S.K. contrived the study and designed the experiments.
F.I., M.M.J.W., J.W.S.,  R.N.T., G.J.L., P.C.D.W.H., R.S.E.,  A.J.P.E.V., M.J.v.d.B., and M.S. included patients in the different studies.
S.R.v.d.V., F.I., M.M.J.W., G.K., R.G., J.W.S., R.N.T., G.J.L., P.C.D.W.H., R.S.E., P.J.F., H.J.D., A.J.P.E.V., M.J.v.d.B., and M.S. collected the data.
S.R.v.d.V. carried out the experiments.
S.R.v.d.V., F.I., M.S., and S.K. interpreted the results.
S.R.v.d.V., F.I., M.J.v.d.B., M.S., and S.K. created the initial draft of the paper.
M.M.J.W., G.K., R.G., J.W.S., R.N.T., G.J.L., P.C.D.W.H., R.S.E., P.J.F., H.J.D., A.J.P.E.V., W.J.N., and M.J.v.d.B. revised the paper.

\bibliographystyle{unsrtnat}
\bibliography{bibliography}
\end{document}